\documentclass[12pt,preprint]{aastex}

\usepackage{graphicx}
\usepackage{multirow}
\usepackage{rotating}
\usepackage{amsmath, amsthm, amssymb}
\usepackage{color}
\usepackage{url}

\shorttitle{Benchmark Tests for MCMC Eclipse Fitting}
\shortauthors{Rogers et al.}

\begin{document}

\title{Benchmark Tests for Markov Chain Monte Carlo Fitting of 
Exoplanet Eclipse Observations
}

\author{Justin Rogers \altaffilmark{1,2},
Mercedes L\'opez-Morales \altaffilmark{3,2},
D\'aniel Apai \altaffilmark{4},
Elisabeth Adams \altaffilmark{3}
}

\affil{e-mail: rogers@pha.jhu.edu}

\altaffiltext{1}{Johns Hopkins University, Department of Physics and Astronomy, 366 Bloomberg Center, 3400 N. Charles Street, Baltimore, MD 21218, USA}
\altaffiltext{2}{Carnegie Institution of Washington, Department of Terrestrial Magnetism, 5241 Broad Branch Rd. NW, Washington D.C., 20015-1305, USA}
\altaffiltext{3}{Harvard-Smithsonian Center for Astrophysics, 60 Garden St., Cambridge, MA 02138, USA}
\altaffiltext{4}{Department of Astronomy, The University of Arizona, and Department of Planetary 
Sciences, The University of Arizona, Tucson, AZ, USA}

\date{Received \today}

\begin{abstract}
Ground-based observations of exoplanet eclipses provide important clues to the planets' atmospheric physics, yet systematics in light curve analyses are not fully understood. It is unknown if measurements suggesting near-infrared flux densities brighter than models predict are real, or artifacts of the analysis processes. We created a large suite of model light curves, using both synthetic and real noise, and tested the common process of light curve modeling and parameter optimization with a Markov Chain Monte Carlo (MCMC) algorithm. With synthetic white-noise models, we find that input eclipse signals are generally recovered within 10\% accuracy for eclipse depths greater than the noise amplitude, and to smaller depths for higher sampling rates and longer baselines. Red-noise models see greater discrepancies between input and measured eclipse signals, often biased in one direction. Finally, we find that in real data, systematic biases result even with a complex model to account for trends, and significant false eclipse signals may appear in a non-Gaussian distribution. To quantify the bias and validate an eclipse measurement, we compare both the planet-hosting star and several of its neighbors to a separately-chosen control sample of field stars. Re-examining the Rogers et al. (2009) Ks-band measurement of CoRoT-1b finds an eclipse $3190^{+370}_{-440}$ ppm deep centered at $\phi_{me}$=$0.50418^{+0.00197}_{-0.00203}$. Finally, we provide and recommend the use of selected datasets we generated as a benchmark test for eclipse modeling and analysis routines, and propose criteria to verify eclipse detections.

\end{abstract}


\section{Introduction}\label{sec:intro}

Eclipses of transiting exoplanets, detected as a dip in flux while the planet passes
behind the star, provide a wealth of information about the planets' atmospheric properties.
The depth of an eclipse in a given wavelength passband provides information about the planet's 
thermal or reflected emission, or both, relative to the flux of the host star.
Therefore, an eclipse detection, even in a single passband, can constrain the day-side 
temperature and atmospheric energy circulation of the planet at a given atmospheric height. 
A measurement at optical wavelengths can also place an upper limit on the planet's albedo 
(see e.g.~\citealt{2007ApJ...667L.191L}). 
Multiple photometric detections at different wavelengths, or equivalently a spectrum, 
can further identify the albedo and energy transfer, and
provide information about the chemical composition of the planets' atmospheres, 
their vertical temperature profiles, and the presence or absence of thermal inversion 
layers (e.g.~\citealt{2008ApJ...678.1436B,2008ApJ...678.1419F}).

To date, more than a dozen exoplanets have published eclipse detections. Most of those 
are from the {\it Spitzer} {\it Space} {\it Telescope}, including eclipses in the four 
mid-infrared bands of the IRAC instrument at 3.6, 4.5, 5.6 and 8 $\mu$m 
(e.g.~\citealt{2005ApJ...626..523C,2008ApJ...684.1427M,2009ApJ...690..822K}), 
at 16 $\mu$m with IRS 
(e.g.~\citealt{2006ApJ...644..560D,2008Natur.456..767G}), and at 24 $\mu$m with MIPS
(e.g.~\citealt{2005Natur.434..740D,2008ApJ...686.1341C}).  
Recently, optical and near-IR eclipse detections have 
also been made with the {\it Hubble} {\it Space} {\it Telescope} \citep{2009ApJ...690L.114S}, 
the {\it CoRoT} and {\it Kepler} missions 
\citep{2009A&A...501L..23A,2009Natur.459..543S,2009Sci...325..709B,2012AJ....143...39C}, and also
with ground-based telescopes 
\citep{2009A&A...493L..31S, 
2009A&A...493L..35D, 
2009A&A...506..359G, 
2009ApJ...707.1707R, 
2010A&A...513L...3A, 
2010MNRAS.404L.114G, 
2010ApJ...716L..36L, 
2010ApJ...717.1084C, 
2010ApJ...718..920C, 
2011MNRAS.416.2096S, 
2011AJ....141...30C, 
2011A&A...530A...5C, 
2011A&A...528A..49D, 
Zhao2012, 
2012ApJ...748L...8Z, 
2012ApJ...760..140C}. 

The optical and near-IR detections are of particular interest, since they can be 
made from the ground, can provide information about the planetary albedos (optical 
observations), and because they are closest to 
the blackbody emission peaks of hot jupiter exoplanets. 
This last factor is important because it allows an estimation 
of the bulk of the energy emitted by those planets.
One feature observed in the ground-based detections to date 
is that several eclipses appear too deep (i.e.~the planet is too bright) in the near-IR 
(around 2.0 $\mu$m), which 
suggests that the planets are hotter than 
predicted by standard models. 
This has been seen for CoRoT-1b in the Êeclipse observations by two independent teams at 
similar wavelengths, i.e.~in Ks-band (2.2 $\mu$m) \citep{2009ApJ...707.1707R}
and in NB2090 band (2.09 $\mu$m) \citep{2009A&A...506..359G}, for WASP-19b in NB2090 band
\citep{2010MNRAS.404L.114G}, for WASP-12b in Ks-band \citep{2011AJ....141...30C}, 
for HAT-P-1b in Ks-band \citep{2011A&A...528A..49D}, 
and most recently for WASP-3b in Ks-band \citep{2012ApJ...748L...8Z}.
These results may be attributed to a real physical effect, suggesting alternate atmospheric
condition scenarios -- for example, unexpected chemical abundances or non-LTE conditions. 
However, because the signal-to-noise ratios are low, any bias or systematic effects in 
the analytical methods used to detect and model the eclipses could dramatically change the 
eclipse measurements and conclusions.

The most common methods used to measure physical characteristics from transit and eclipse light curves
involve constructing a parameterized model of the light curve and its systematic effects, and then 
using a routine -- typically a Markov Chain Monte Carlo (MCMC) algorithm -- to derive the 
best-fitting set of parameters according to a simple criterion such as the $\chi^2$ or Bayesian 
Information Criterion value (e.g. \citealt{2007Natur.447..691H}). 
Some alternatives exist; for example, SysRem \citep{2005MNRAS.356.1466T}, wavelet-based algorithms 
\citep{2009ApJ...704...51C}, non-parametric algorithms \citep{2011arXiv1106.1989W}, and 
Gaussian processes \citep{Gibson2012}. These are discussed in Section~\ref{ss:dmtech}.

In this work, however, we focus on what remain the most common methods for measuring the signals and 
potential systematics associated with exoplanet eclipse observations 
from the ground, i.e.~MCMC algorithms, 
and present benchmark data sets to test the effectiveness of these methods. 
In Section~\ref{sec:methods} we describe the characteristics of a typical ground-based eclipse observation
and the data modeling and signal extraction process. 
In Section~\ref{sec:implementation} we describe our model implementation to test potential 
systematic effects associated with the steps described in Section~\ref{sec:methods}. 
Section~\ref{sec:synthdata} presents the results of the tests for synthetic data, both with and without
red noise, and in Section~\ref{sec:realdata} we show the results of those same tests using real datasets. 
In Section~\ref{sec:corot1} we discuss the measurement of real eclipses and our 
re-evaluation of the depth and uncertainty of the published 
Ks-band eclipse of CoRoT--1b. 
Our conclusions and the implications for
future exoplanet eclipse searches are summarized in Section~\ref{sec:discussion}.
Appendix~\ref{sec:benchmark} provides a collection of 
benchmark data sets that can be shared for testing and verifying eclipse modeling and analysis processes.


\section{Secondary Eclipse Observations and Detection Methods}\label{sec:methods}

The methods described in this section focus on ground-based observations of secondary eclipses, 
which are becoming the primary means of exploring 
the optical and near-IR portions of exoplanetary emission spectra, 
but are currently made at only modest confidence levels (typically less than 10 $\sigma$). 
However, this same description may also apply to primary transit observations 
and to observations from space, if the significance levels of the signals are low.

Detections of eclipses in the optical and near-IR are presently limited to ``very hot Jupiters'' (VHJs), 
i.e.~Jupiter-size exoplanets which reside very close to their host stars (less than 3-day orbits) 
and are expected to be tidally locked, with typical atmospheric day-side temperatures 
around 1500--2500 K. 
The large radii and hot day-sides make these VHJs the most favorable exoplanets for detection. 
However, even in these cases, the eclipse signals of VHJs 
are typically very small ($<$ 0.1\% in the optical and 
$<$ 0.5\% in the near-IR), comparable to the scale of the best point-to-point 
differential flux uncertainties currently achievable by ground-based detectors at those wavelengths. 

The precision with which secondary eclipses can be detected is typically limited by correlated, 
or \emph{``red''} noise, caused mainly by systematic trends related to changing atmospheric and 
instrumental parameters \citep{2006MNRAS.373..231P}.  Although noise levels should 
be improved by collecting a large number of data points during one or multiple eclipse epochs, 
this is the case if the only noise present is random (Poisson or white noise). 
In practice, the red noise limits the noise reduction, 
which in turn limits the confidence of the eclipse measurement.
This explains all of the published ground-based detections being at the modest 
3-10$\sigma$ confidence level, except for the 
$K_S$-band detection of WASP-12b by \citet{2011AJ....141...30C} ---
by observing the target for 6.2 hours with no dithering and 22 stable reference stars, this eclipse was 
measured to a remarkable 25-$\sigma$ level.
New techniques for detrending and removal of systematics have significantly 
lowered the amplitude of residual red noise with respect to the unbinned dispersion due to Poisson noise, but 
some level of residuals 
are still typical, as seen in most of the published detection papers. 
Even if the instrumental and other systematics are well-accounted for, intrinsic stellar noise itself can 
contribute above the Poisson level \citep{Gilliland2011}.

\subsection{Typical Observations and Dataset Characteristics}\label{ss:typical}

A typical exoplanet eclipse observation consists of a series of short exposures -- a ``light curve'' --
in a single night over several hours (enough to capture the entire eclipse and suitable baseline). 
The most critical attributes determining 
whether the eclipse signal can be extracted 
are the depth of the signal (i.e.~the planet-to-star flux ratio) and the noise 
levels (both red and white components).  
The chief goals are to minimize the point-to-point 
flux dispersion of the light curve and to obtain as many data points as possible, in order to best 
detect a difference between the in- and out-of-eclipse flux levels. 
These goals are complicated, however, as each characteristic of the observation set -- comparison stars, airmass, 
dithering strategy, phase coverage, sampling rate  -- 
contributes to the level and structure of the noise.
There are a number of observational observational techniques employed to optimize the light curve for detection, 
whether from a space-borne dedicated instrument 
(see e.g.~\citealt{Gilliland2011} for the \emph{Kepler} process), 
or using existing ground-based facilities.

Most VHJ eclipses last between 1.5 and 2.5 hours, the time it takes for the planet to pass behind 
the star from the observer's viewpoint. 
Table~\ref{t:exopx} lists the parameters that determine this duration for 
CoRoT-1b \citep{Barge2008,Bean2009}, a VHJ that we have 
targeted for secondary eclipse detection, and a 
synthetic planet that we discuss in Section~\ref{sec:synthdata}.  
$T_{14}$ is the total time from the beginning of the eclipse's ingress (first contact) 
to the end of its egress (fourth contact), i.e.~the ``Full Duration'' of the eclipse; 
$T_{23}$ is the ``Flat-Bottom Duration'' -- the time from second to third contact 
during which that the planet is entirely 
eclipsed by the star (i.e. when the light curve is flat and at minimum flux).

While it is possible to measure the depth of an eclipse 
with only partial coverage (e.g.~the WASP-12b J-band measurement by \citealt{2011AJ....141...30C}), 
the optimal scenario is 
to observe a continuous window including the entire eclipse and some out-of-eclipse baseline 
both before and after.
If the total number of baseline observations is less than the number of observations during 
full eclipse, the confidence level of the eclipse detection will be limited by the baseline's noise 
statistics.  Obtaining a baseline at least as long as the eclipse duration will provide a better 
opportunity to make a detection, with the noise statistics only limited by the number of in-eclipse points.

Obtaining many data points requires a sampling rate as fast as possible.  There is a competing 
interest, however, in obtaining enough counts per frame to limit the photon noise in the 
target and comparison stars.  An established strategy is to set the exposures to be just long enough 
that the dispersion is dominated by correlated noise rather than photon statistics.  
For ground-based observations, this typically means obtaining at least 10$^6$ integrated photons 
per frame (in both the target and comparisons) to limit the photon noise to 10$^{-3}$, 
approximately equal to the contribution level of the red noise, readout errors, flat-fielding errors, etc.  
Longer exposures, limited by the same systematic noise factors, will not substantially improve the 
photometric precision.
For near-IR observations, particularly K-band, there are 
also sky background concerns to consider -- long exposures will saturate, and the faster sky 
variability is another motivation to have a fast sampling rate.  To avoid saturation but 
still obtain enough photon counts, it is sometimes necessary to defocus the star images and spread the 
signal over more pixels on the chip. This allows the collection of more photons and can also 
reduce the sensitivity to small shifts in the stars' locations on the chip, but the greater number 
of pixels used will also contribute more noise.
Modern instruments can employ other methods of improving the observing 
cadence, such as the use of faster-readout subframes or routines designed to optimize the duty cycle. 
For example, the eclipse of OGLE-TR-56b \citep{2009A&A...493L..31S} was 
detected with an E2V camera featuring 
frame-transfer technology allowing almost no readout time and near 100\% duty cycle, and the 
WASP-4b detection \citep{2011arXiv1104.0041C} used a datacube mode 
that can store 250 images at a time with no readout overhead until the end. 
Taking all of this into consideration, the ideal sampling rate is set 
anywhere from one point every couple of minutes to as many as 100 points per 
minute, depending on the telescope's aperture and the apparent brightness of the host and 
comparison stars.

One of the most important systematic noise minimization strategies 
is to have at least one nearby star suitable to be used as a comparison 
for differential photometry. 
An ideal comparison star should be non-variable and similar to the target in brightness 
(to achieve good photon noise in the same exposure time as the target, without saturating) 
and color (to limit differential atmospheric extinction). 
Better still is to have many comparison stars, although this is 
limited by the instrument's field of view and the brightness of the target.  
Measuring the target system's flux relative to the comparison stars, 
rather than the absolute flux count of the target, substantially reduces the 
atmospheric effects, including airmass and thin clouds.  
To minimize chromatic effects, it is also desirable to limit the observations 
to an airmass less than $\sim$1.6-1.7.

To avoid flat-fielding and chip sensitivity errors, it is usually best to keep the stars occupying the same 
pixels throughout the observations.  In the near-IR, noise from the rapidly-varying sky 
background can sometimes be more effectively removed by frequently moving the target and 
comparison stars to different pixels (dithering), and then 
subtracting the images from one another.  
Recently, however, this strategy has often been seen as counter-productive, as systematics from pixel variations 
outweigh the sky background problem.  Keeping the stars in fixed positions for both optical and near-IR 
observations appears to be the best strategy \citep{2011AJ....141...30C}.

\subsection{Data Modeling Techniques}\label{ss:dmtech}

Even with all the noise-minimization 
techniques described in the previous section, a significant amount of red noise systematics 
will likely remain in the observed light curve. 
The next (and most difficult) step is to model 
and remove these systematic trends; otherwise they can 
confound the extraction of the eclipse signal.

The amount of correlated noise in a data series can be 
estimated by calculating the 
dispersion $\sigma_N$ when binning the residuals (i.e.~the difference between the observed data 
and the best-fit model) into $N$ points.  
For an entirely random noise distribution (white noise), $\sigma_N$ will decrease as $N^{-1/2}$, 
but in the presence of red noise it instead follows:
\begin{equation}
\sigma_N^2 = \frac{\sigma_w^2}{N} + \sigma_r^2
\label{e:rednoise}
\end{equation}
where $\sigma_w$ and $\sigma_r$ are the amplitudes of white and red noise, respectively. 
Thus, the contribution of the white noise can be reduced by binning over larger samples, but the red 
noise cannot. 
Over an eclipse light curve, the red noise often contributes 20 to 50\% as much noise as the unbinned white noise level 
before detrending \citep{2006MNRAS.373..231P}. 
The white and red noise amplitudes can be estimated for any data series by finding the 
$\sigma_w$ and $\sigma_r$ values that best suit this equation. 

The red noise is not completely unexplained, however. 
In practice, the detected flux is often found 
to correlate with various effects of the atmosphere and instrument 
(see e.g.~\citealt{2007Natur.447..691H}).  
The observed light curve can be thought of as the product of three components: 
the flux reduction by the actual eclipse, the intrinsic random white noise (photon noise, read noise, etc.), 
and the red noise resulting from systematic correlations.
The goal is to remove the systematic trends and recover 
the basic parameters that describe the shape of the eclipse signal: 
the eclipse depth $D_e$, 
which represents the emission of the planet in the observed passband, 
and the mid-eclipse phase $\phi_{me}$, which provides information about 
the eccentricity.

There are several approaches to model and remove these systematics, 
and several sophisticated algorithms have been proposed and tested. 
For example, for fields with many stars, algorithms like SysRem \citep{2005MNRAS.356.1466T} can
automatically measure trends and remove them from the light curve to reduce the 
red noise (e.g.~\citealt{2009A&A...493L..31S}). 
\citet{2009ApJ...704...51C} 
introduced a wavelet-based method to estimate the parameters. In it, the noise is treated as the sum 
of time-correlated and uncorrelated components, and the likelihood function is calculated with a 
fast wavelet transform. 
\citet{2011arXiv1106.1989W} has
proposed a non-parametric algorithm to 
blindly de-convolve all non-Gaussian signals from the eclipse light curve. 
This method is best suited for simultaneous observations at different wavelengths (e.g.~binned 
spectroscopic observations); most photometric detections are based on a single light curve, or
curves of eclipses on different nights.
Both of these alternative methods 
preferentially remove the correlated noise without making use 
of the measured atmospheric and instrumental quantities that are often used to seek correlations and 
systematic effects.
In another method, \citet{Gibson2012} have used Gaussian processes for regression, which define a distribution 
over function space, and maintain some advantages of both the linear parameter model and the Waldmann 
algorithm. Their technique makes use of additional measured parameters, but then finds the distributions of 
systematics in a non-parametric way.

Most common, however, is to simply look for correlations between observed flux and measured 
atmospheric or instrumental parameters, and remove them to reveal the signal. 
The most basic method is to use only the out-of-eclipse baseline 
points of the light curves, i.e.~the portions that are expected to be constant in time, to 
manually look for and remove trends one at a time, beginning with the most strongly correlated.  
For example, if the baseline flux appears to correlate strongly with airmass, one can fit a linear trend 
between airmass and the differential flux in the baseline, and then remove this trend from the entire light curve. 
This process can be repeated iteratively with other parameters until the trends remaining in the baseline are minimized.  
Since these fits are done using only the out-of-eclipse portions of the light curve, systematics 
in the in-eclipse data are removed by interpolation of the out-of-eclipse fits.
Many of the ground-based eclipse detections 
(e.g.~\citealt{2009ApJ...707.1707R,2011AJ....141...30C}) 
have been made in this manner.

Some weaknesses of looking for trends only in the out-of-eclipse portions of 
the light curve are that it requires a prior guess of which points are not in the eclipse, 
and it does not make use of the large amount of in-eclipse data, in which other trends could be 
present.  The manual removal of one correlation at a time can also be problematic in cases 
where there is not a dominant trend, as the removal of a second trend can reintroduce 
correlations with the first parameter.

Arguably a better approach is to consider the effects of the correlation trends 
and the shape of the eclipse signal simultaneously together in a model.
This can be done by constructing 
a correlation function $C_q(q)$ with each independent variable $q$ one might pick out from 
the observations 
(e.g.~airmass, position on the detector, sky brightness, exposure time, shape of the images, etc.), 
and then combining those functions with the eclipse light curve $F_{ecl}$ to form 
a more detailed model for the observed flux:
\begin{equation}
F_{obs} = F_{ecl} \prod_q{C_q(q)}
\label{fulllcmodel}
\end{equation}
At first-order one may use linear $C_q$ functions, although in fact the best correlations may be non-linear. 
This has been demonstrated to be a 
useful technique, simultaneously accounting for the physical effect of the system and the 
atmospheric and instrumental effects \citep{2007Natur.447..691H}.

A challenging aspect of this method, however, is the need to test a large number of free parameters, 
without initially knowing which parameters are the most essential. 
Testing a comprehensive grid of possible combinations of these parameters can be 
time-prohibitive;
instead, a method is needed to efficiently find the set of parameters, with uncertainties, that 
best models the observed light curve.  A few different algorithms are used to achieve this, 
but the most common is the MCMC, which we 
discuss in Section~\ref{ss:mcmc}.

We consider this last approach to be the most robust; therefore it is the one we have focused on in the 
rest of this work. Our implementation is discussed in detail in Section~\ref{sec:implementation}.

\section{Model Implementation and Systematic Effects Tests}\label{sec:implementation}

To test the analysis process which extracts eclipse detections 
from photometric light curves, we set out to first generate a variety of test datasets 
(discussed in Sections \ref{sec:synthdata} and \ref{sec:realdata}), 
and then 
evaluate them using our own adaptation of a multi-parameter model of the eclipse 
plus correlation trends, and MCMC routines to find the best-fit set of parameters.
The goal is to determine 
how well this type of model and the MCMC process measure
an input eclipse of known depth and phase placement.  
If the process is working effectively, the best-fit solution should recover the 
same correlation and eclipse signal that was input. 
Under a range of noise levels, systematics, and other conditions, how close does the MCMC come 
to recovering a signal of a given depth?  
At what noise limits do we no longer detect the input signal?  Are there any systematic errors in the measurement?
In order to answer these key questions, we required our own implementation of an observed light curve 
model and a process for fitting the model with the best parameters.

\subsection{The Light Curve Model with Systematic Trends}\label{ss:lcmodel}

We began constructing our model of the light curve of a secondary eclipse of a typical VHJ with a 
simple eclipse curve - the light curve that the system's flux would produce in the absence of any noise. 
This shape is determined by a number of 
star-planet system parameters, 
generally constrained by transit and radial velocity observations to better precision than 
achievable from eclipse observations: stellar radius $R_*$, planet radius $R_P$, 
orbital period $P$, semi-major axis $a$, inclination $i$, and eccentricity $e$. 
We selected for these quantities values comparable to those of CoRoT-1b and other representative VHJs, 
as given in Table~\ref{t:exopx}. 

%
%
%
%

Similarly to the transit models in \citet{2002ApJ...580L.171M}, 
the eclipse curve was generated by simulating the motion of the planet relative to the star, 
and calculating from this the fraction of the planet's surface area eclipsed $P_{ecl}$ 
as a function of orbital phase $\phi$.  The eclipse window for a transiting VHJ is
typically a small enough portion of the period that orbital eccentricity affects only 
$\phi_{me}$, the phase at which the middle of the eclipse is centered, which is directly 
related to $e\cos{\omega}$, where $\omega$ is the argument of the periastron.
We began by calculating 
the function $r(\phi)$, the separation between star and planet disk centers in the plane of 
the sky, in units of $R_*$:
\begin{equation}
r(\phi)=\frac{a}{R_*}[1-\sin^2{i}\cos^2{(\phi-\phi_{me})}]^{1/2},
\label{rofphi}
\end{equation}
and from there we found 
the portion of the planet that is eclipsed as a function of $r$:
\begin{equation}
P_{ecl}(r)=
\begin{cases} 0 & r \ge 1+ p  \\ 
 \frac{1}{\pi p^2} (\theta_1-\sin{\theta_1}\cos{\theta_1})+
 \frac{1}{\pi}(\theta_2-\sin{\theta_2}\cos{\theta_2}) & 1-p < r < 1+p   \\ 
1 & r \le 1 - p \\  
\end{cases}
\label{pecl}
\end{equation}
where $p$ is the unitless ratio $R_P/R_*$, and the angles $\theta_1$ and $\theta_2$ are defined by: 
\begin{equation}
\cos{\theta_1}=\frac{1+r^2-p^2}{2r}; 
\cos{\theta_2}=\frac{r^2+p^2-1}{2rp}.
\label{e:costh12}
\end{equation}

Because it is the planet and not the star that is eclipsed, stellar limb darkening does not affect the 
eclipse shape, and the portion of the light curve when the planet is fully obscured will be flat.  
The flux decrement, or depth $D_e$, of the eclipse depends on the brightness (thermal and reflected emission) of the planet. 
For our purposes, we considered the planet to be seen as a uniformly bright disk, allowing us to 
model the eclipse shape as 
\begin{equation}
F_{ecl}(\phi) = F_b * \{1 + D_e * [1-P_{ecl}(\phi,\phi_{me})]\}
\label{secondary}
\end{equation}
where $F_b$ is a normalizing baseline flux parameter.  
$P_{ecl}(\phi,\phi_{me})$ was found using 
Equations \ref{rofphi}, \ref{pecl}, and \ref{e:costh12}, the 
mid-eclipse phase $\phi_{me}$, and the other fixed parameters, 
and then used with $D_e$ to model 
the expected flux from the star and planet throughout the eclipse (flux of 1 in full 
eclipse, 1+$D_e$ when the planet is fully visible, normalized to the brightness of the star alone).
Thus, with all the other star and planet parameters held constant, the model for the eclipse 
shape depends only on $F_b$, $D_e$, and $\phi_{me}$.

Next, in order to model the systematic effects, we allowed possible relationships 
between the detected flux and various independent variables measured in the dataset, 
constructing for each variable a simple linear correlation function:
\begin{equation}
C_q(q) = 1 + m_q q
\label{cq}
\end{equation}
where $m_q$ is the correlation slope factor between a given independent 
variable $q$ and the flux. Each 
linear function could in principle also have an intercept $b_q$, but these can
easily be combined into the 
overall baseline $F_b$, so they are not needed in each correction curve.

Note that this linear function is merely the simplest possible correlation one could devise. 
For measurements such as pointing/pixel effects, it is an approximation for effects that could be 
nonlinear. Thus, this model can be expanded to include polynomial or other correlation functions, 
e.g.~a quadratic function $C_q(q) = 1 + m_q q + m_{q2} q^2$.
Later, we explored this in our ``real data'' tests (Section~\ref{ss:nlc}).

We considered several variables that might be available and relevant: 
the airmass ($z$), the full-width half-maximum size of the star images ($w$), their 
x- and y- positions on the chip ($x$, $y$), sky brightness ($s$), and 
time (or, equivalently, orbital phase $\phi$).  
Correlations with other characteristics can of course also be modeled as needed, but the
model parameters we selected for this work are these, also listed in Table~\ref{t:mplist}.

Some of these 
($\phi$, $z$) are universal in each dataset, but 
the rest ($m_w$, $m_x$, $m_y$, $m_s$) 
are measured separately for each star in the field. As long as a variable exhibits the 
same behavior over time for each star (e.g.~if one star moves three pixels in the x-direction and its 
FWHM increases by 10\%, the same happens to all other stars in the field), a single set of values can 
be used to determine that trends with that variable. 
Even though each star may be affected differently by changes in one parameter, the differential flux 
can be modeled with a single correlation. 
However, if the behavior of the same variable differs greatly from star to star, 
it may be better to use a different set of measurements, and model parameters, for each star 
(e.g.~if there is a different sky brightness behavior in different parts of the field, use $s_0$, 
$s_1$, etc.~rather than just one $s$ -- and increase the number of corresponding slope factors $m_s$).

In addition, as discussed in Section~\ref{ss:typical}, a single dataset may comprise 
observations from multiple offset / dither positions, each of which can exhibit different trends. 
For example, $x$, $y$, and $w$ each affect the distribution of photons across different pixels on the detector, 
so their effect on the measured flux will not be the same for different offset positions. 
The two primary parameters ($D_e$, $\phi_{me}$) in Table~\ref{t:mplist} are 
global to the entire dataset, defining the shape and 
placement of the eclipse in the light curve, but 
the rest can have different behavior at different offset positions; thus, we use 
a separate set of these parameters to apply to the points at each offset position.

All the correlation functions for the independent variables are then multiplied by the noiseless secondary 
eclipse light curve (Equation \ref{secondary}) to form a more accurate model for the 
observed flux of the target planet-star system as a function of orbital phase $\phi$, 
and the adjustable model parameters:
\begin{equation}
F_{mod} = F_{ecl}(\phi,F_b,D_e,\phi_{me}) \prod_q{C_q(q)}.
\label{fullmodel}
\end{equation}

By finding the parameters that produce a single best-fit match of the model in 
Equation~\ref{fullmodel} 
to the observed 
light curve, we derive the key 
values we are seeking: the eclipse's depth and central phase.

\subsection{The MCMC}\label{ss:mcmc}

With differently-behaving parameters, multiple stars and offset positions, the number of parameters 
required to produce the full light curve model can begin to grow rapidly, and 
testing even a sparse grid of possible combinations can become prohibitive. 
To more efficiently sample the parameter space and find the best 
solution with uncertainties, we implemented two separate MCMC algorithms, 
each developed independently to ensure that we were not seeing the 
effects of a problematic procedure.

The first MCMC process, which we called ``MCMC-A'', begins by generating a set of 
initial guesses for all the parameters, 
and in each step randomly 
selects one parameter to perturb by a small amount, or ``jump.''  
Each step generates a model light curve for this set 
of parameters including the jump, and calculates an information criterion, or 
``fitness'', for example, the $\chi^2$ value between that 
model and the observed points (see Section~\ref{ss:ic}).  The new fitness is compared 
to the fitness at the previous step, and then the jump is either accepted or rejected, following the 
Metropolis-Hastings algorithm:
If the new fitness is lower (i.e.~better) than the previous one, the jump is accepted, and 
the chain proceeds to the next step with a set of parameter values 
including that new jumped-to value; 
if the new fitness is worse than the previous one, the jump may still be accepted, with a probability 
$P_{jump} = e^{-\Delta\chi^2}$, where $\Delta\chi^2$ is the difference in fitness caused by this jump.
The second option prevents the solution from remaining in a local minimum in the parameter space.  
If the jump is rejected, the pre-jump parameters are carried over 
to the next step.  This process of random jumps being accepted or rejected is repeated up to 
four million times, resulting in  
a distribution of values for each parameter from all the steps of the MCMC.  
To remove biases from the starting position, we discard the first 20\% of the steps, 
and then examine the remaining distribution for each parameter.  
These values are then binned into a histogram, 
and the midpoint of the most heavily-populated bin (i.e.~the mode average) is taken to be the 
best-fit value of that parameter, with the errors set such that the central 68\% of the points 
fall within the 1-$\sigma$ limits.

For every eclipse we tested, MCMC-A first ran two short (10,000-step) MCMC trials to 
adjust the jump size for each parameter, 
i.e.~the amplitude of the random-normal distribution from which each jump is selected. 
If a jump size were set too small, the distribution would be too diffuse; if the jumps
were too large, the chain would not jump enough.
After setting jump sizes for each parameter that would result in 30 - 40\% of jumps being accepted, we 
ran fifteen separate MCMC's -- three starting from each of five 
different selected initial parameter guesses, in order to test the effects that the 
starting position would have on the chain.
We repeated each test with two different information criteria used for the Metropolis-Hastings test: 
$\chi^2$ and the Bayesian Information Criterion (described in Section~\ref{ss:ic}).
All of the data described in the following sections (synthetic and real datasets) were 
analyzed with this MCMC-A implementation.  

We also tested a representative sample of light curves 
with a second, independent MCMC analysis process by a
different team member, using the model described in
\citet{Carter2011}. This analysis, which we called MCMC-B, 
used a Gibbs sampler to vary a random parameter for each step, before also using
the Metropolis-Hastings jump acceptance criterion. Before each long
chain, a sequence of shorter sequences was run to tune the step sizes
for each parameter to between 20-40\%. For each test, one chain of
1,000,000 values was run, with the first 5\% discarded, and the median 
and standard deviation of the truncated chain for each parameter are reported 
as the best-fit values and errors.

\begin{figure}
{\centering
  \includegraphics[height=0.9 \textwidth,angle=0]{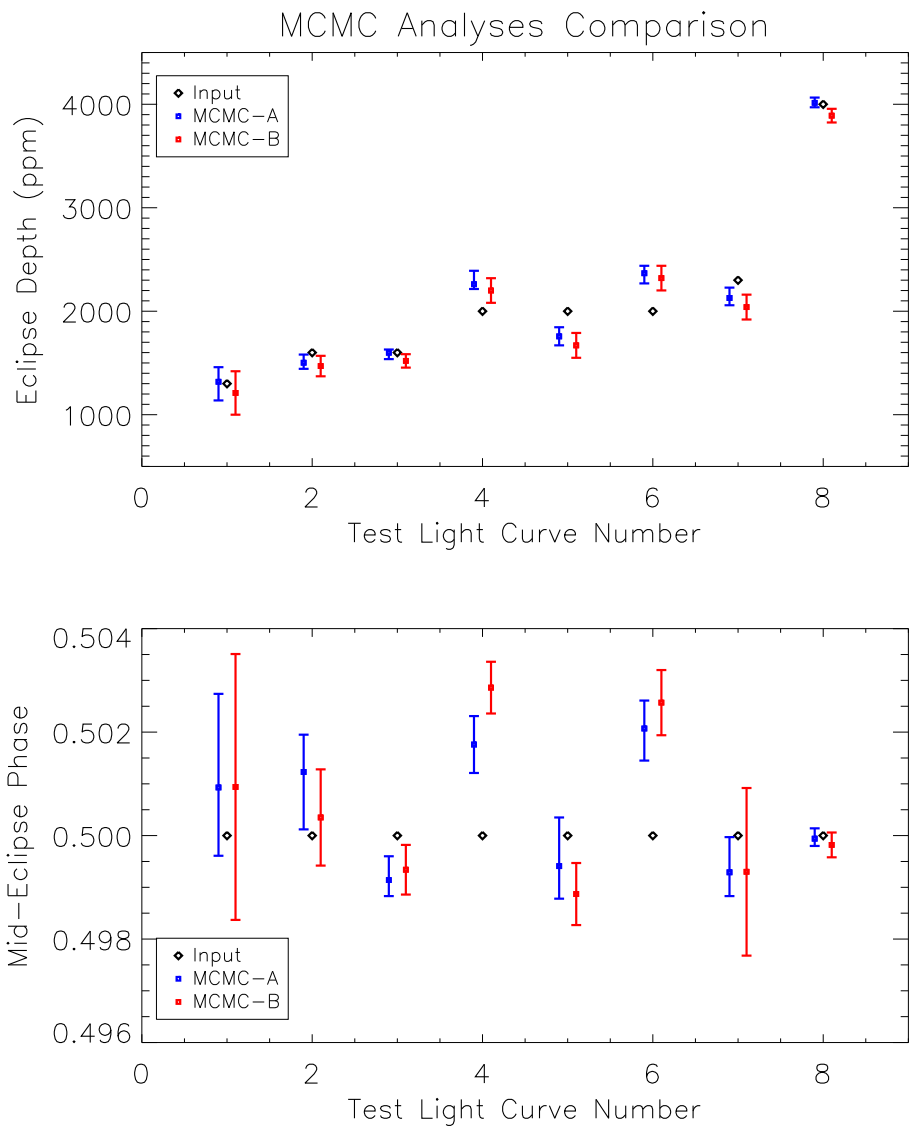}}
\caption[]{
A comparison of the results from each of the two independently developed 
MCMC analysis processes we tested 
on eight selected synthetic light curves.  Identical sets of light curves, covering a range 
of eclipse depth, baseline length, sampling rate, and red noise amplitude and pattern, were 
run through each analysis, in order to compare the results to each other and to the input values.  
The x-positions of each result are offset for purposes of clarity.
The results from both MCMCs were consistent with each other, and 
are also shown in Table~\ref{t:jrvsea} along with a description of the light curves. 
\label{f:jrvseafig}
}
\end{figure}

When run independently on the same datasets, the two MCMC processes 
produced similar results within their reported uncertainties.  
Table~\ref{t:jrvsea} compares the results between the two processes on a 
sample of eight selected synthetic light curves. The first three columns describe the 
characteristics of each dataset: the amplitude $\sigma_r$ and pattern of the 
red noise used in the light curve (see Section~\ref{ss:rn} and Figure~\ref{f:rnmodels}), 
the length of the observing window, and the sampling rate of the 
observations. Columns 4 through 6 list the eclipse depth that is put into the light curve 
initially, and recovered by each MCMC process, where ``Fit (A)'' and ``Fit (B)'' 
correspond to MCMC-A and MCMC-B respectively. 
Columns 7 through 9 show the inputs and results for the mid-eclipse phase.
All eight tests used the $\chi^2$ information criterion, and had an input 
white noise amplitude $\sigma_w$ set to 1000 ppm.
The best-fit solutions for eclipse depth $D_e$ and mid-eclipse phase are also shown in 
Figure~\ref{f:jrvseafig}. 
From these tests we concluded that both independently-developed MCMCs 
produce consistent results. Thus confirming the reliability of each, we proceeded to 
use MCMC-A in the remainder of this work.


\subsection{The Information Criterion}\label{ss:ic}

When including parameters in the model, 
we do not initially know which are relevant and which will 
simply complicate the process without
helping to model the systematics; for example, one light curve may have a strong 
dependence on the position of the stars on the chip, but little to no effect from the 
changing airmass. 
In such a case, allowing a dependence on airmass only increases the uncertainty 
on the other parameters without improving the fit. 
Thus, a quantitative measure is needed to describe both 
the goodness of fit and the usefulness of the free parameters.  
For that purpose, in addition to the basic $\chi^2$ value, 
we also tested an alternative, the Bayesian Information Criterion (BIC), 
that takes into account the number of parameters used.  
The BIC is defined as:
\begin{equation}
BIC = \chi^2 + k \ln{n}
\label{e:bic}
\end{equation}
where $n$ is the number of observed points and $k$ is the number of nonzero parameters.  
The purpose of the BIC is to introduce a preference for simpler models over more 
complex models that may be ``overfitting'' the data with more free parameters than needed. 
A complex model (i.e.~one with a higher number of nonzero parameters $k$) 
will still be preferred in some cases, but only if it significantly improves the fit to overcome 
the greater $k \ln{n}$ term. 
For example, removing one parameter from the model (i.e. setting it to zero) 
reduces the BIC value by $ln(n)$, so if doing so does not raise the 
$\chi^2$ value by more than that amount, the solution with one fewer parameter is preferred.

The MCMC will not naturally jump a parameter to exactly zero, so in order to implement the BIC, 
we set a threshold range for each parameter such that jumps to values near zero would be 
set to zero to minimize the BIC. We found that using a threshold larger than a few times 
a parameter's adjusted jump size would raise rather than lower the BIC. Thus, we set the 
threshold for each parameter equal to the jump size 
(previously adjusted as described in Section~\ref{ss:mcmc}).


\section{Synthetic Datasets}\label{sec:synthdata}

As discussed in Section~\ref{ss:typical}, the 
recovery of a secondary eclipse signal 
by the modeling and analysis process depends on the depth 
of the eclipse and the noise levels, as well as 
other aspects of the data set such as phase coverage and sampling rate.  
Before beginning on real data, which has unknown noise patterns, 
we wanted to test our method
on completely synthetic datasets with known and controlled noise sources.  
To do this, and investigate 
how each characteristic might affect the process's results, we 
created a wide variety of synthetic eclipse light curves, and then ran them 
through the MCMC-A process to see what depths and phases 
would be measured.
Each test dataset consisted of a secondary eclipse curve with a controlled eclipse
depth $D_e$, and mid-eclipse phase $\phi_{me}$, combined with a noise profile.

\subsection{Light Curve Construction}\label{ss:lcconstruction}
We began by creating a model ``Exoplanet X'' similar to many detected VHJs.  
The planet and star radii, orbital period, semimajor axis, and inclination, which constrain the duration 
and shape of the eclipses, are listed in Table~\ref{t:exopx}.  
The period, 25 hours, was selected for an easy 
conversion between time and phase -- every hour of observation corresponds to a span of 0.04 in phase.  
This configuration results in a flat-bottom portion of the eclipse ($T_{23}$) spanning 94 minutes, with ingress 
and egress lasting 18 minutes each, for a full eclipse duration ($T_{14}$) of 130 minutes.

For most of the trials, we used an observation window beginning two hours before 
and ending two hours after the mid-eclipse time (240 minutes total, extending from phase 0.42 to 0.58).
This provided 55 minutes of baseline on each side 
of the eclipse, giving a total baseline of 110 minutes, which is slightly longer 
than the time spent in the flat-bottom portion of the eclipse shape. 
We also tested runs with shorter and longer baselines (15, 94, and 188 minutes on each side; 
total durations of 160, 318, and 506 minutes), and a 240-minute series with the 
mid-eclipse phase offset to 0.475, to provide a baseline that was short (17.5 minutes) 
on one side and long on the other.  

We then set up an array of observation points across the selected window with a given sampling rate, and 
a small random error (normally distributed; $\sigma$=0.1 s) added to the timing of each point to 
simulate imperfect data spacing.  We tested sampling rates of 25, 75, and 250 points per hour 
(100, 300, and 1000 data points, respectively, in the four-hour observing window).
Table~\ref{t:synthparams} lists all the values we adjusted and tested in this section.

Next, using these phase points and the planet model, we built the (noiseless) secondary 
eclipse shape with the same function (Equation \ref{secondary}) used for that part of the 
light curve model.  
Given all the fixed star and planet values, and a set baseline flux $F_b=1$, the only adjustable 
parameters that would adjust the noiseless eclipse shape were $D_e$ and $\phi_{me}$. 

Finally, we generated a noise profile to add to the eclipse curve, with either white noise only 
(Section~\ref{ss:wn}) or a combination of white and red noise (Section~\ref{ss:rn}).
An example noiseless eclipse shape is shown 
in panel (a) of Figure~\ref{f:allkinds}, along with different noise models,
constructed light curves, and their best-fit model solutions.

\subsection{MCMC Evaluation}\label{ss:mcmceval}

As described in Section~\ref{ss:mcmc}, the MCMC-A process ran fifteen 
Markov chains for each light curve evaluated, 
starting from five distinct sets of model parameter values.
Only the first four parameters ($D_e$, $\phi_{me}$, $F_b$, and $m_{\phi}$) from
Table~\ref{t:mplist} were used in fitting the synthetic eclipses, as 
values for the other measured atmospheric and 
detector effects did not exist. 
For real datasets (Section~\ref{sec:realdata}), we would include all 
of the parameters.

As an example, 
Table~\ref{t:exresults15} shows the results for a complete set of fifteen trials 
on one light curve:
75 data points per hour in a four-hour observing window, with a 
synthetic Exoplanet X eclipse of depth $D_e$=4000 parts per million (ppm), 
centered at phase $\phi_{me}$=0.500, 
and red noise (Pattern 3; see Section~\ref{ss:rn} and Figure~\ref{f:rnmodels}) 
equal in magnitude to the white noise (1000 ppm), 
and run to optimize the $\chi^2$ information criterion.  
All of these aspects are discussed in Section~\ref{ss:rn}, and 
this is also the light curve shown in panel (d) of Figure~\ref{f:allkinds}.

The trials are divided into five subsets of three trials, each one set starting from different 
initial guesses for $D_e$ and $\phi_{me}$. The trials are identified in the first column by their subset's 
starting point and then the trial number within that subset, e.g.~1-2 for the second trial 
at the first starting position.
The second and third columns provide the initial guesses for each subset.
Columns four through six show the best-fit results for the eclipse depth, mid-eclipse phase, and 
baseline flux (phase slope $m_\phi$ is not shown). The final two columns show two measures of 
the model's fitness -- the $\chi^2$ value and the root mean square of the residuals.

The results among each set of trials were generally in good agreement and
independent of the point from which the chain was started, though 
occasionally the series from one starting point would find a less optimal solution, 
which was reflected by a higher dispersion of the residuals between the observed and model points.
For example, in the set of trials shown in Table~\ref{t:exresults15}, 
the recovered values from four of the five starting 
points were in agreement with each other 
within their errors, and had very similar $\chi^2$ and residual RMS values, suggesting that the space 
they populated represented the global best fit. 
Meanwhile, the dispersions of the residuals from starting point 4 were much worse, 
indicating that the MCMC found only a local minimum. 
Of the fifteen trials, we adopted the solution with the lowest residual dispersion to provide the 
model light curve and best-fit parameters. 
The last row of the table lists the key results from this example's best fit (trial 5-1):  
eclipse depth (adjusted for the baseline), and mid-eclipse phase. 
This solution differs from the input depth by 360 ppm (7.3$\sigma$, but only 9\% of the input depth), 
and the input mid-eclipse phase by 0.00084 (3.2$\sigma$), equivalent to 76 s for the orbit of Exoplanet X. 
Both of these values suggest that the errors may be underestimated.

\subsection{White Noise Only}\label{ss:wn}

In the simplest scenario, we generated normally-distributed white noise to add to the 
noiseless eclipse curves.  
Because the detectability would be dependent on the depth-to-noise ratio, we chose 
to keep the white noise amplitude $\sigma_w$ (i.e.~the RMS of the normal distribution) 
fixed at 0.1\%, or 1000 ppm, and test different eclipse depths,   
beginning at 5000 ppm (five times the white noise level) and continuing down to 50 ppm, 
as well as different values for the other attributes of the light curves (see Table~\ref{t:mplist}). 
Panel (b) in Figure~\ref{f:allkinds} shows a 
white noise pattern and the combined light curve (eclipse plus noise) with its best-fit model. 

\begin{figure}
{\centering
  \includegraphics[width=0.9 \textwidth,angle=0]{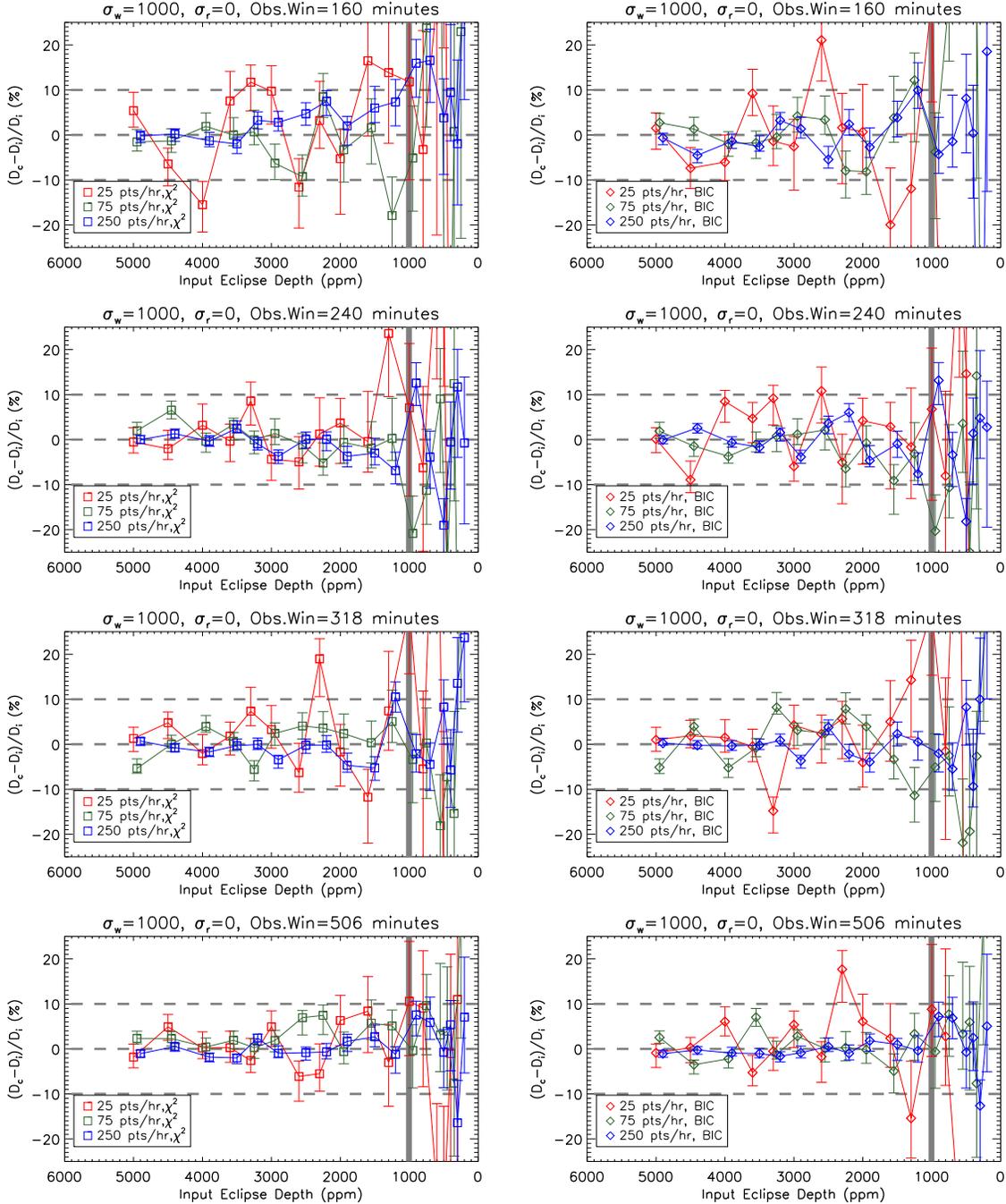}}
\caption[]{The behavior of the best-fit recovered eclipse signals as a function of 
input depth, for the tests with white noise only. The gray vertical bar shows 
the white noise level of $\sigma_w$=1000 ppm.
The accuracy is expressed as a percent difference between the input 
and recovered depth.  
Each row of plots represents a different baseline length, and the colors 
show the different sampling rates.  
The plots in the left column use the $\chi^2$ 
information criterion, while those in the right column use the BIC.
The x-positions of the symbols are slightly offset in order to distinguish the data points and error bars.
Following each series of points from left to right, the critical depth is defined as the point at which
the depth percent errors leave the $\pm$10\% range for the second time.
\label{f:R0_pd}
}
\end{figure}

Figure~\ref{f:R0_pd} shows the difference between the eclipse depths that were input into 
the tests and those that were measured by the analysis, as a percentage of the input depths. 
We tested the range of eclipse depths for four different observing window lengths, and using both 
the $\chi^2$ and Bayesian information criteria. Each of these sets is shown in a separate plot, 
with the three different sampling rates in different colors.
At the highest input depths, both the eclipse depth and mid-eclipse phase that were 
recovered were in close agreement with the values from the input models, as expected; 
as the model depths were decreased, these values began to disagree by greater amounts.
By looking at each series starting with the deepest input and moving to the shallower ones, 
we were able to examine under 
what conditions the signals continued to be recovered successfully.

To quantify the results, we calculated the percent difference between the recovered and input 
values at each eclipse depth, beginning with the deepest and looked for a
point below which we could no 
longer recover the eclipse signal reliably (for our purposes, we referred to a 
recovered eclipse depth within 10\% of the input depth as ``successful''). 
We then defined a ``critical depth'' $D_{e,crit}$ 
as the input depth of the last successful 
recovery before the \emph{second} time 
the difference was greater than 10\% (to avoid being confounded by one outlier point).
We found that the mid-eclipse phases were also recovered 
accurately for tests above this threshold -- 63\% contained 
the input $\phi_{me}$ within 1$\sigma$ of the calculated values and errors; 91\% within 2$\sigma$ --
but diverged as the input depths became shallower. 
Above the critical depth, we calculated a ``dispersion'' as
the average of the absolute values of the errors after discarding the highest and lowest values, 
and considered whether the deviations were systematic or random in nature.
Figure~\ref{f:R0_pd} illustrates the accuracy of the recovered eclipse depths for all the 
synthetic, white noise only tests, and the critical depths and average errors for every 
combination of observing window, sampling rate, and information criterion are listed in 
the columns in Table~\ref{t:cdd} labeled ``$\sigma_r$=0''.

Typically, the signals were recovered in agreement with the inputs 
as long as the depth was at or above the white noise level (1000 ppm).  
The only instances where this was not the case were at the lowest sampling rate 
(25 points per hour), where there were fewer points in the light curve; however, 
even in these instances the results were not off by more than about 15\% until we passed 1000 ppm. 
At depths below that level, the process was much less likely to recover the input depth, although the 
observing window and sampling rate did affect this.  

For the standard 240-minute observing 
block, $D_{e,crit}$ was at or near the white noise level (800 -- 1000 ppm) when 
using both the $\chi^2$ and BIC criteria. 
We also found that, while the sampling rate had little effect on the critical depth reached, 
it did affect the dispersion of the results for the points above $D_{e,crit}$.  
As we increased the sampling rate, the recovered depths clustered closer to the input values. 
This showed, as expected, that a faster sampling rate at the same noise level will return 
a more accurate measurement.

In the case of the shorter (160-minute) observing windows, eclipse depths were recovered less 
accurately, and the critical depth was reached sooner.
For the longer windows (particularly the longest, 506 minutes) and 
at the faster sampling rates, accurate 
depths were recovered from eclipses with signals as low as 250 to 300 ppm, i.e.~a factor 
of three to four times shallower than the white noise level.   
This demonstrated that baselines that significantly exceed the duration of the eclipse are preferable.

In all cases tested, there were no systematic biases toward measuring the eclipses to be 
deeper or shallower eclipses than the inputs; rather, the errors appeared random.

%
%
%
%

\subsection{Red Noise}\label{ss:rn}

Next, we created an additional component of 
the noise model to simulate the effects of red noise.
\citet{2009ApJ...704...51C} took a multi-step approach to synthesizing red noise:
first, they established the covariance conditions for wavelet and scaling 
coefficients for the case of combined white and red noise. 
Then they drew 
a sequence of 1024 independent random variables, set such that their 
wavelet and scaling coefficients would obey these conditions, 
and finally, took the inverse Fast Wavelet Transform of the sequence to 
generate a red noise signal.

We constructed our initial models in a simpler fashion, 
by adding together four sine waves with periods equal to 
2, 1, 0.5, and 0.25 times the four-hour observation window 
(i.e.~$P_n$= 480, 240, 120, or 60 minutes):  
\begin{equation}
R(\phi) = \sum_{n=1}^4 A_n \sin{(\phi/P_n+\omega_n)}
\label{e:rnform}
\end{equation}
This model has 
eight adjustable parameters (amplitude $A_n$ and 
phase angle $\omega_n$ for each component).  
We chose sets of parameter values to create three distinct 
red noise models, shown as Patterns 1, 2, and 3  
in Figure~\ref{f:rnmodels}, 
and measured the RMS of the points in each pattern.
Then to use a given red-noise pattern, we scaled the amplitude to set the RMS  
to a desired red noise level $\sigma_r$.
We began with a red noise amplitude of $\sigma_r$ = 200 ppm 
(20\% of $\sigma_w$, a typical level for the red noise found in differential light curves), 
but also tested higher red noise levels of 500 ppm and 1000 ppm. 

While we anticipated that our models using a few frequencies would produce similar effects as typical observed 
noise, real systematics can include power at all frequencies, and increasing for longer wavelengths. 
Thus, in addition to the initial patterns, we generated Brownian noise ($1/f^2$) models by 
integrating series of Gaussian white noise. 
Two such patterns are shown in Figure~\ref{f:rnmodels} as Patterns 4 and 5. These were scaled 
and added into the light curves the same way as the sinusoidal patterns.

For each test, we then combined the red-noise pattern with the randomized white noise 
pattern (always kept at $\sigma_w$ = 1000 ppm) 
and eclipse shape, and ran the resulting light curve 
through the MCMC and modeling process as before in the white noise tests. 
An example light curve (noise and eclipse signal) for Pattern 3, $\sigma_r$=1000 ppm, and 
a synthetic eclipse of $D_e$ = 4000 ppm 
is shown in panel (c) in Figure~\ref{f:allkinds}.

\begin{figure}
{\centering
  \includegraphics[width=0.85 \textwidth,angle=0]{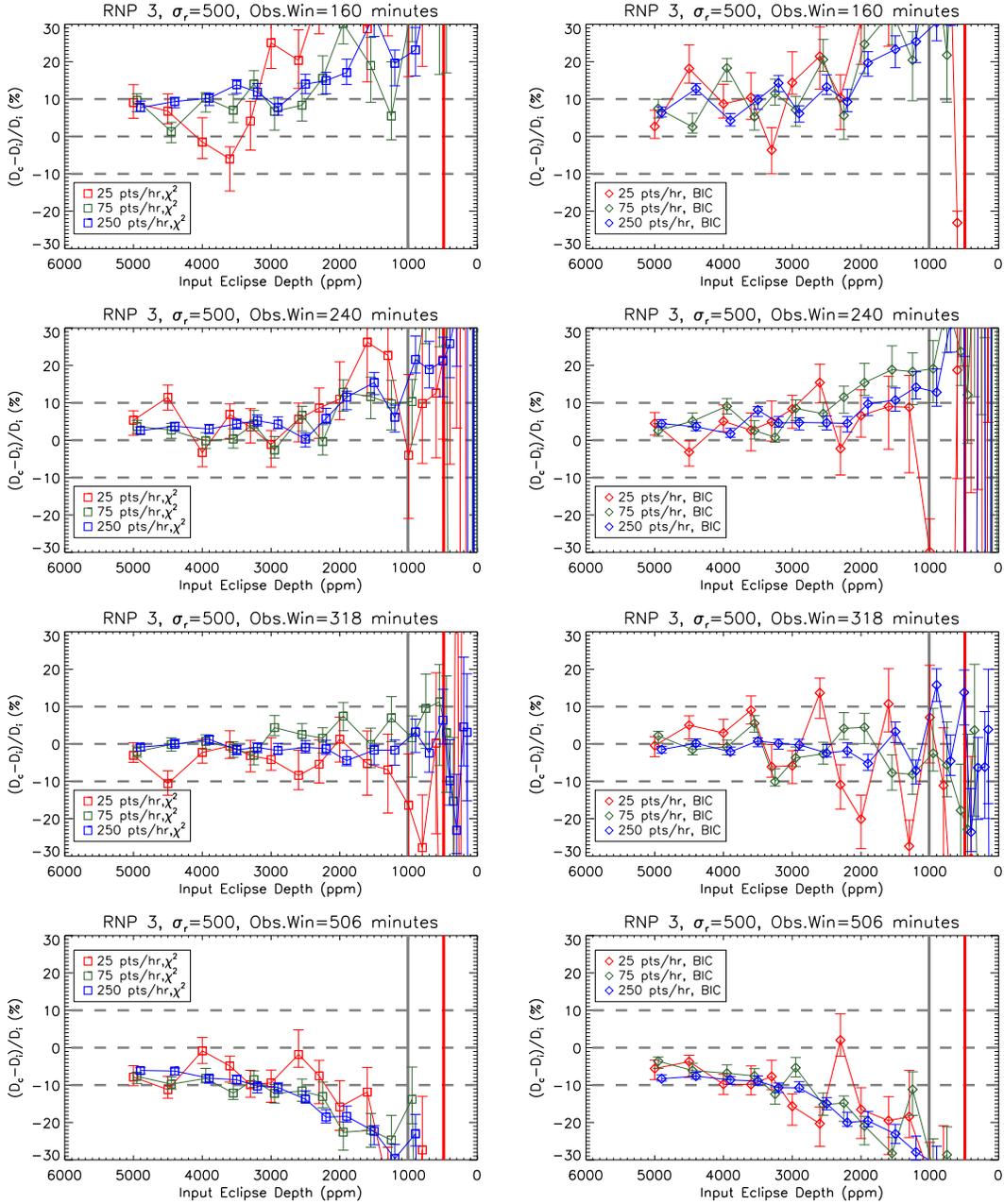}}
\caption[]{The behavior of the best-fit recovered eclipse signals as a function of 
input depth, for the tests with white noise and red noise (Pattern 3) added. 
The amplitudes of these noise patterns ($\sigma_w$=1000 ppm, $\sigma_r$=500 ppm) 
are shown by the vertical gray and red bars.
The accuracy is expressed as a percent difference between the input 
and recovered depth.  
Each row represents a different baseline length, and the colors red, green, and blue 
represent the different sampling rates. 
The plots in the left column use the $\chi^2$ 
information criterion, while those in the right column use the BIC.
The x-positions of the symbols are slightly offset in order to distinguish the data points and error bars.
\label{f:R3_500_pd}
}
\end{figure}

Figure~\ref{f:R3_500_pd} shows the recovery results for the light curves with a red noise component (Pattern 3)
of amplitude 500 ppm, i.e.~50\% of the white noise level, added to the raw eclipse shape and white noise.  
These are subdivided by observing window length, information criterion, and sampling rate. 
In all of these tests, we saw little difference between the $\chi^2$ and BIC results, due to the fact that 
these models had fewer free parameters than the real data tests with instrumental and atmospheric measurements. 
In Section~\ref{ss:rdresults}, we will note that this is no longer the case.

We calculated critical depths and error dispersions in the same manner as for the white noise tests 
(Section~\ref{ss:wn}); these results for every observing window / sampling rate / information criterion 
combination are also given in Table~\ref{t:cdd} in the columns marked ``$\sigma_r$=500 ppm''.
As in the white-noise case, the accuracy of both the depth and 
mid-eclipse phase measurements diverged as the input depths became smaller, 
but this began to happen at depths greater than the white noise level, 
or even the combined point-to-point uncertainty from the white and red noise together.
The dispersions above the critical depths were greater as well, and we saw stronger systematic 
errors than in the white noise tests.

In the series of 240-minute tests for all three red noise levels, the recovered eclipse 
depths were within $\sim$10\% of the input depths down to near the white noise level, 
yet the best-fit models consistently measured eclipse signals slightly deeper than the inputs, 
as seen in the second row of Figure~\ref{f:R3_500_pd}. 
The critical depths for this baseline ranged from 600 ppm (60\% of the white noise level)
to as large as 3300 ppm (over three times the white noise amplitude) for $\sigma_r$=200 ppm and 500 ppm. 
For inputs above the critical depths, both information criteria methods resulted in eclipse measurements 3 to 6\% 
deeper than the input depths.
For the $\sigma_r$=1000 ppm series, the critical depths were all 2300 ppm or higher, and the dispersion was 
greater (5 to 9\%).
With the shorter 160-minute baseline, there was a stronger bias towards deeper detections than 
the inputs, but the critical depths were much higher as well (2000 to 4000 ppm).

Baselines of 318 minutes showed improvement in $D_{e,crit}$ and dispersion, but 
when the baseline was extended to 506 minutes, the accuracy became systematically worse again, 
with each series tending to measure the eclipse signals around 5\% shallower than the inputs for 
$\sigma_r$=200 ppm, and even more so for the higher levels of red noise.  
Above the critical depths, these deviations were greater than the error bars on the depths measured 
by the MCMC, particularly for the cases with faster sampling rates.

The strongest biases were seen in the tests with the most red noise ($\sigma_r$=1000, equal to $\sigma_w$). 
Especially in the 160- and 240-minute baselines, the recovered eclipses were offset from the input values 
by more than 10\% even for the highest input depths (see again Table~\ref{t:cdd}).
These trends illustrate the red noise floor explained by Equation~\ref{e:rednoise} -- increasing the 
number of points reduces the error from the white noise, but not from the red noise.
As expected, the critical depths and dispersions increased with higher red noise amplitudes. 

Examining the shape of red noise Pattern 3 in Figure~\ref{f:rnmodels} indicated why the 
different baselines showed different systematic biases.
For the longest baseline (506 minutes, extending from phase 0.33 to 0.67), the  
full red noise pattern was lower on both ends and high in the middle, which counteracted the inserted 
eclipse shape. When the ends were clipped to a shorter phase coverage, that influence was 
removed, and when clipped even further (to 240 or 160 minutes), 
the dip at phase 0.51-0.52 became a signal that was added to the eclipse depth.
With Pattern 1, the longest-baseline set showed a bias recovering deeper eclipses than input; 
with Pattern 2 there was little bias either way. 
These also suggest that with a long observing window it may be prudent to attempt fits 
with only the necessary portion of it, since extended baselines can have trends that confound 
the eclipse measurement. 

The effects of the Brownian noise patterns (4 and 5) on one series (four-hour observing window, sampling 
rate of 250 points per hour) are shown in Figure~\ref{f:rn45}: for both $\chi^2$ and 
BIC, Pattern 5 behaves similarly to the simpler Pattern 3, though with a slightly higher positive bias at most 
depths. Pattern 4 acts in the opposite direction with similar amplitudes, resulting in slightly lesser 
measured eclipse depths. They did not show qualitatively different results than the sinusoidal noise patterns.

A key finding was that the resulting uncertainties from the MCMC process far underpredicted the actual 
offsets between input and measured eclipse depths. For example, in the Pattern 4, $\sigma_r$=500 ppm, 
$\chi^2$ series, the measured depths differed from the input depths by 2\% to 10\% above $D_{e,crit}$, but 
the estimated errors from the MCMC were between 0.7\% and 3\%, a factor of 2 - 4 smaller.

These results showed that the unknown structure of the red noise, even in the baseline portions 
of the light curve, can introduce in addition to the random error an additional systematic bias affecting all
the measurements on a series of tests with a fixed noise pattern in a certain way. However, in practice on 
a real dataset, this bias is only a once-observed effect, which amounts to a greater uncertainty on the 
actual depth of the eclipse event. This was reflected by the disagreements between the input and calculated 
eclipse depths being greater than the error bars found by the MCMC process.


\section{Real Data Systematics}\label{sec:realdata}

After a thorough exploration of synthetic light curves, we began to test the process on 
real data, which would have measurable white and red noise levels, but unknown 
systematics and trends. We took some photometry from a previously observed exoplanet 
field to be treated as the noise and, as in the previous section, added a range of simulated 
eclipse signals to see which could be recovered.

\subsection{The Dataset and Light Curves}\label{ss:rdataset}
The dataset we used for these tests was the photometry from the 
CoRoT--1b field, taken in Ks-band with APO/NICFPS, on UT 15 January, 2009, and published in 
\citet{2009ApJ...707.1707R} (hereafter R09).  
This set included 758 images alternating between two offset positions 15 arcseconds apart.  
However, finding the differential photometric precision to be substantially worse at high airmass, 
we removed the points at $z>$1.8, as well as any points that deviated in flux by more than 3$\sigma$ 
from their neighbors within 20 minutes, leaving 691 data points.
In addition to the target, there were four stars of relatively similar magnitude 
(between 0.95 and 1.8 times the target flux) in the same area of the chip, and many more stars 
in the field.
We assigned each star a number from 0 through 13, with 0 being the target, CoRoT-1, 
and 1 through 4 the bright nearby reference stars used in R09. 
The remaining images spanned 293 minutes, with a small gap (8.8 minutes) in the coverage near 
phase 0.55, resulting in a cadence of approximately 145 points per hour. 
The observing window and sampling rate for our real dataset tests were fixed by 
those characteristics of the dataset itself, unlike in the case of the synthetic tests.

To produce light curves for use with our synthetic eclipse tests, we selected 
bright, stable field stars one at a time to treat as though they were the target, 
and performed differential photometry with respect to other nearby stars in the field, using 
the comparison stars which produced the lowest flux RMS in the differential 
light curves. To avoid issues with the actual eclipse in the data, we excluded the 
main target star CoRoT-1 from this process.
For each light curve, we then measured the amount of red and white noise as the 
values for $\sigma_w$ and $\sigma_r$ that best fit 
the binned dispersions to Equation \ref{e:rednoise}, finding that 
both were greater than in the synthetic noise tests. 
For example, the noise levels for star 1's differential light curve  
were found to be $\sigma_w = 4580$ ppm and $\sigma_r = 1540$ ppm. 
We then also made the same measurements using just the out-of-eclipse baseline points, since 
for a measurement on the target light curve we would want to use only these points to judge 
the noise structure. In most cases the levels were very similar to those in the full light curve.
We repeated this for several stars in the field, numbered 1, 2, 3, 4, 8, and 13, in order 
to test the systematics of each. 
For stars 1 and 4, the best comparisons and the measured red 
and white noise for each (in ppm) light curve are given in the third and fourth columns of 
Table~\ref{t:realresults} (identified as ``1-a'' and ``4-a'').

\begin{figure}
{\centering
  \includegraphics[width=0.9 \textwidth,angle=0]{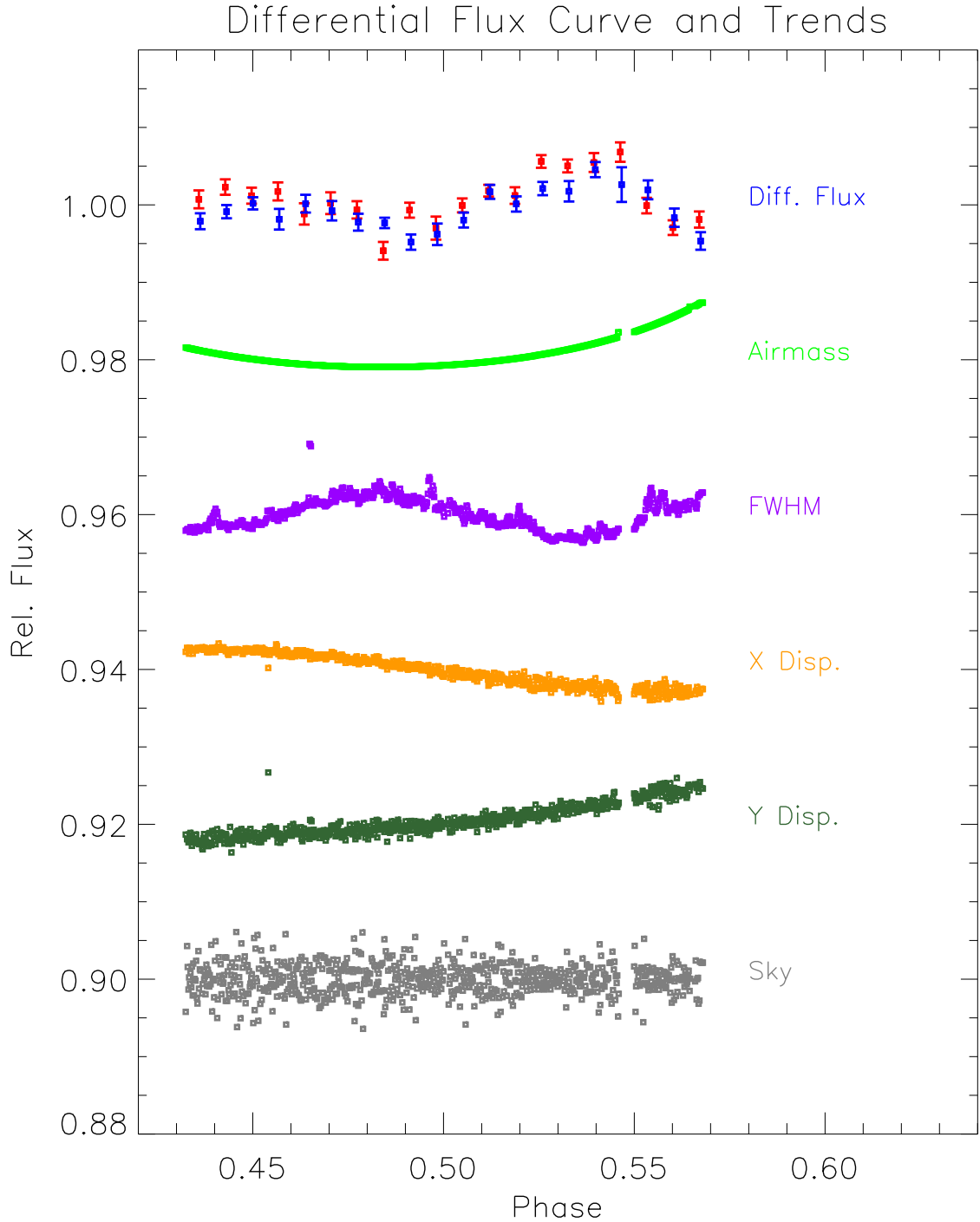}}
\caption[]{
The light curve and trends from the real dataset (CoRoT-1b field in Ks).  
The top set of points (red and blue; each represents one of the two offset positions)
shows the differential flux between Comparison 
Star 1 and the others in the field, 
to be used as real noise for the tests in Sections \ref{sec:realdata} and \ref{sec:corot1}.  
The points are in 15-minute bins for clarity. 
Below are curves (normalized, not to scale with one another)
showing the other parameters measured to look for correlation.
\label{f:ds1_trends}
}
\end{figure}

The differential flux for light curve 1-a is shown at the top of 
Figure~\ref{f:ds1_trends} in blue and red, 
with each color representing one of the 
two offset positions.  
Below it are the other measured characteristics that we used  
for de-correlation: the airmass ($z$), full width at half-maximum of the star images ($w$), 
displacement of the stars on the chip in the x- and y-directions ($x$,$y$), 
and the sky brightness ($s$). 
In some datasets we have seen these attributes behave differently for each star; 
In this particular dataset, however, 
all of these comparison stars were localized to the same area, 
and we tested that the values behaved similarly for each star.
Therefore, we were able to use single arrays for each variable 
$\phi$, $z$, $w$, $x$, $y$, and $s$, although each variable still required two different 
slope coefficients -- one for the behavior at each offset position. 

In total, the light curve model required 16 parameters (see Table~\ref{t:mplist}: 
$F_b$, $m_{\phi}$, $m_z$, $m_w$, 
$m_x$, $m_y$, and $m_s$ for each of the two offset positions, along with the 
global parameters $D_e$ and $\phi_{me}$.
When using the $\chi^2$ value as the fitness criterion, the models would make use of all 
16 parameters, allowing correlations with 
each variable / offset position combination. 
In the BIC runs, however, we expected that 
most of the best-fit models would use only the most influential parameters, 
with the others fixed to zero for a better BIC value.

\subsection{Real Dataset Results}\label{ss:rdresults}

The resulting light curves from each test star's differential photometry were then used 
as the noise components of new eclipse light curves to be tested. 
As before in Section~\ref{sec:synthdata}, 
we constructed a noiseless eclipse shape for 
each given depth and mid-eclipse phase, with the same star and planet parameters
of the hypothetical Exoplanet X (Table~\ref{t:exopx}).
We then combined 
this signal with the real noise profiles to form light curves to be analyzed by our 
modeling and parameter optimization processes, and then ran the tests for 
a range of eclipse depths, two different mid-eclipse phases (0.5 and 0.475), and both the 
$\chi^2$ and BIC information criteria. While the models with $\chi^2$ optimization used 
all sixteen model parameters, the BIC runs typically chose only nine to twelve, omitting some of 
the correlation slope parameters (see Table~\ref{t:realresults}).

\begin{figure}
{\centering
  \includegraphics[width=0.9 \textwidth,angle=0]{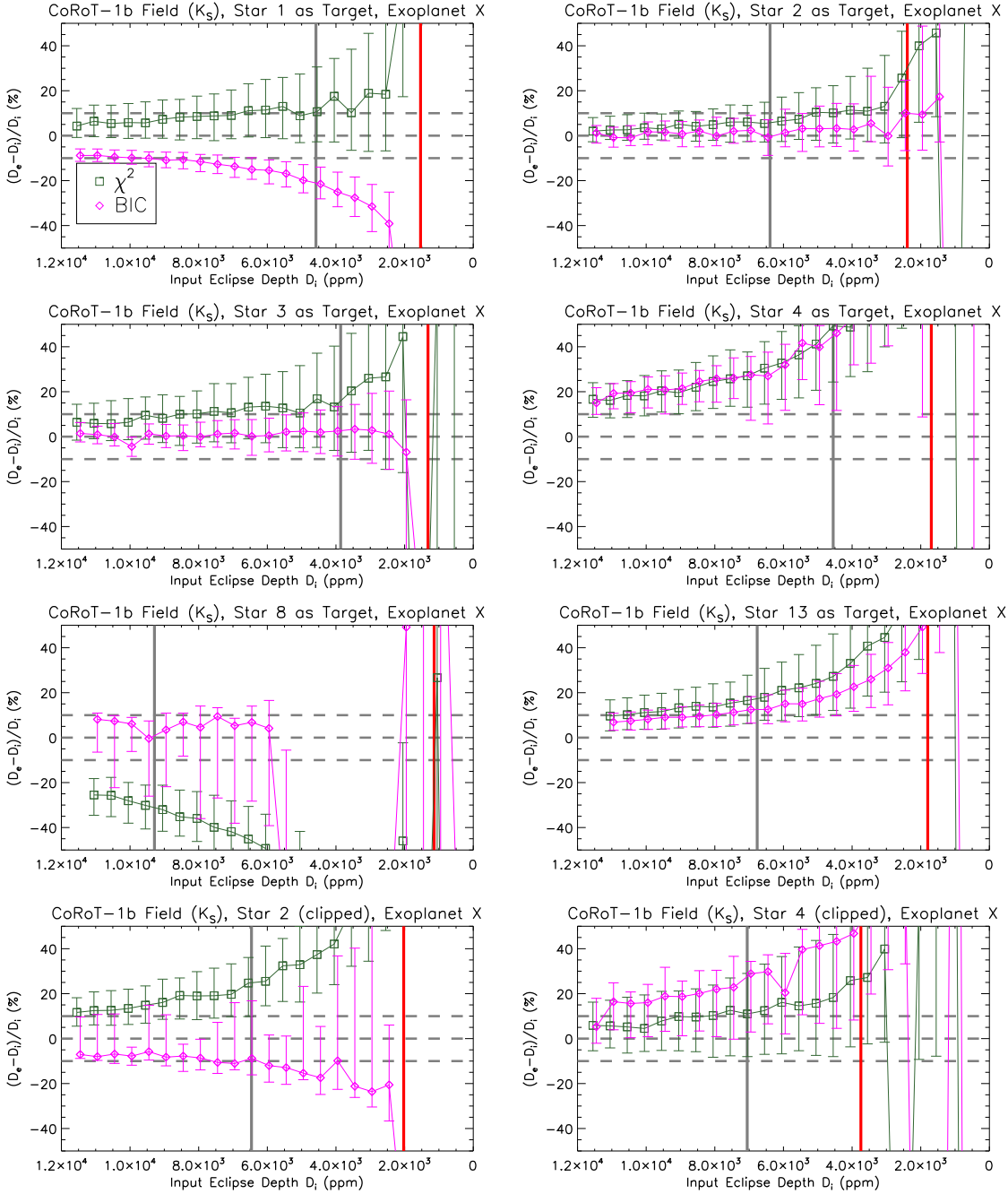}}
\caption[]{
The depth recovery results from the combination of real noise and synthetic eclipses of varying 
input depths. 
The dark green squares represent the results using $\chi^2$, while the pink diamond points 
represent the results using the BIC.  The measured white and red noise 
amplitudes are indicated by the vertical gray and red bars, respectively. 
Each of the top six panels uses a different star as the target, while the 
The bottom row shows results from two light curves -- one from a well-behaved series (Star 2), and one 
from a poorly-behaved series (Star 4) -- clipped to only 50 minutes before and after the eclipse.
\label{f:pd_corot1k_all}
}
\end{figure}

The behavior of the recovered eclipse signals can be seen in Figure~\ref{f:pd_corot1k_all}.
In most cases, as with the synthetic tests, the recovered eclipse parameters were close to the 
input values for large input eclipse depths, but began to deviate more at depths lower than 
the white noise level.  Unlike in the synthetic tests, which had a newly-generated random 
white noise component for each trial, the noise structure was the same in every run for a 
given target star. This generally resulted in calculated eclipse depths consistently higher or 
lower than the input depth, with the systematic bias getting worse as 
the input model depths decreased.

For each test star's series, both the uncertainties on the depth measurements and the 
differences between the measured and input depths were 
substantially higher than in the synthetic-noise tests. This led to higher values for the 
critical depth $D_{e,crit}$, ranging from 2000 to 9500 ppm, and in some cases (e.g.~for star 4) the 
recovered eclipse depths were far outside the 10\% range for all depths tested.
This was partially due to the point-to-point uncertainty being greater, as evidenced 
by the higher measured white and red noise amplitudes, shown in each window of 
Figure~\ref{f:pd_corot1k_all} as vertical gray and red bars. 
Notably, there was far more difference between the $\chi^2$ and BIC results for these series than 
for the synthetic model tests, because there were far more parameters that could be set to zero 
for different preferred solutions.

For the best-behaved test stars (2 and 3), the critical depths occurred at or below the 
white noise amplitude. In these well-behaved series we saw better behavior from the BIC 
than from the $\chi^2$ models.
However, the others (1, 4, 8, and 13) had strong biases from their inputs at 
greater depths, and had mixed results comparing $\chi^2$ and BIC. 
Star 8, for example, behaved well with the BIC to below the level of $\sigma_w$ before 
getting abruptly worse, while for $\chi^2$ it differed by more than 
20\% even from input signals greater than 10$^4$ ppm.

The disagreements between the reported errors and the actual offsets between input and measured 
eclipse depths varied substantially as well. For example, for Star 1 and Star 4 using the BIC, 
the offsets were 3 - 4 times greater than the error bars, whereas for Star 1 using $\chi^2$ and Star 2 
using either criterion, the error bars appeared to be appropriately-sized, containing many of the 
offsets within 1 $\sigma$.

Recalling from the tests in Section~\ref{ss:rn} that different baselines can affect the depth 
measurement, we also tried clipping the baseline on two of the light curves: 
one well-behaved series (star 2), and one poorly-behaved one (star 4).
In addition to running a series with all 
the available points, we ran tests with only the points between phases 0.444 and 0.556 
(572 points over 241 minutes). 
The series for star 2 was made significantly worse by removing the extra baseline, but interestingly, 
the tests for star 4
resulted in values closer to the input depths when 
using $\chi^2$ (though still mostly outside the 10\% error threshold) while the BIC results remained about the same.  
The results for these two clipped light curves 
can be seen in the bottom row of Figure~\ref{f:pd_corot1k_all}.

Even if we account for the different noise sources and consider the eclipse depths relative to the 
noise levels, it is apparent that systematic biases exist, and are different for each choice of target star and the 
amount of baseline. In practice, it appears that each star's light curve has its own systematics 
relative to the others, which can lead to biases in a measured eclipse signal.

\subsection{Non-linear correlations}\label{ss:nlc}

As discussed in Section~\ref{ss:lcmodel}, some of the parameters may have non-linear effects; 
e.g.~although we modeled them as such, there is no 
reason that the relationship between flux systematics and position on the chip, or the size and shape of the star images, 
should behave linearly. To explore this, we added the use of a quadratic fit to the x- and y-displacement parameters, since 
those are values that may most likely have non-linear shapes. This increased the number of model parameters by two 
per offset position, with two coefficients in the $x$ and $y$ models rather than one.

\begin{figure}
{\centering
  \includegraphics[width=0.9 \textwidth,angle=0]{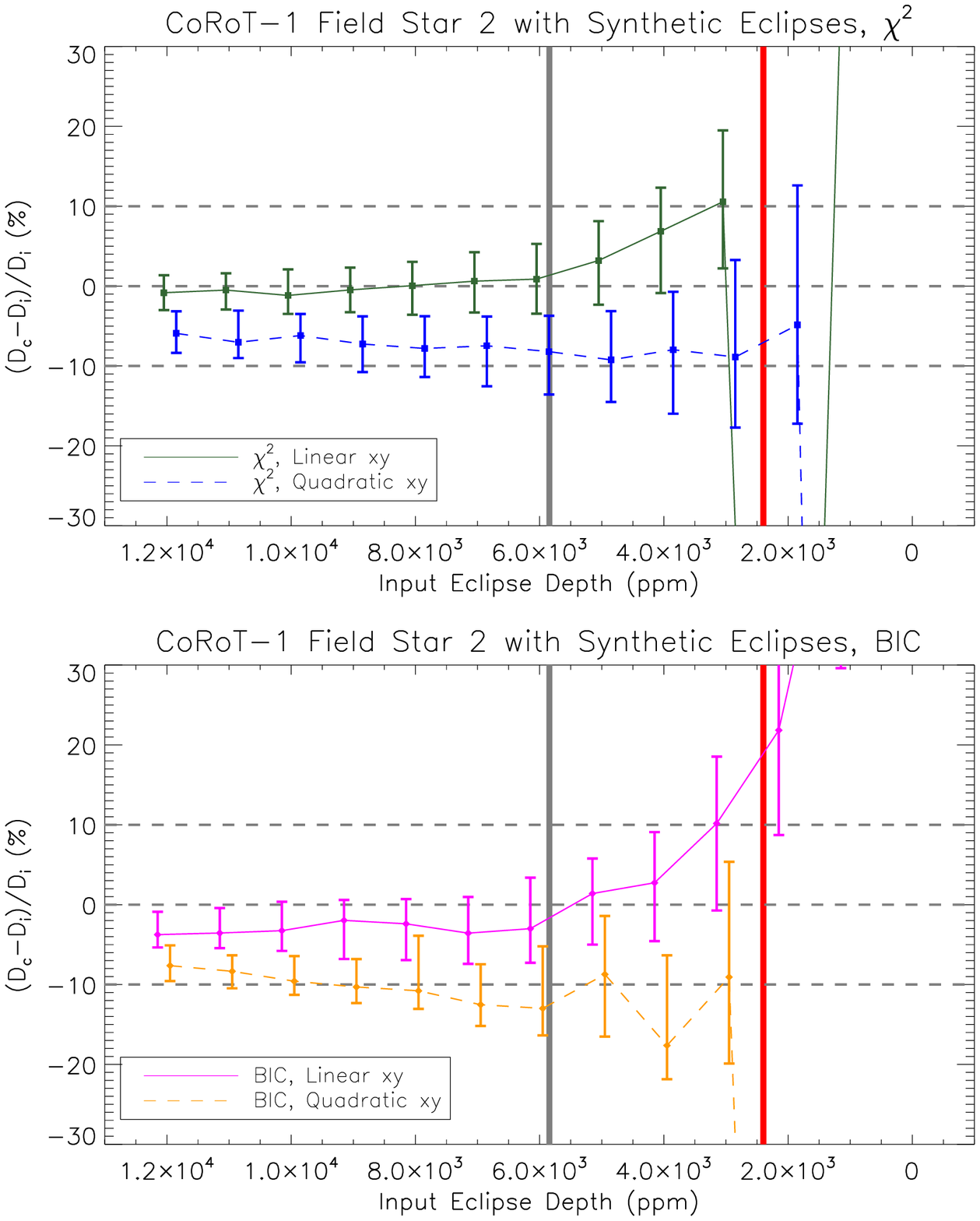}}
\caption[]{
Comparison of the results from using the linear and quadratic models for dependence on 
x- and y-displacement, for both $\chi^2$ (top panel) and BIC (bottom panel). 
Each of these are performed using the Star 2 light curve, with synthetic 
eclipses of varying depths input.
The measured white and red noise 
amplitudes are indicated by the vertical gray and red bars, respectively. 
\label{f:qxy}
}
\end{figure}

Figure~\ref{f:qxy} shows the results in the same format as Figures~\ref{f:R0_pd}, \ref{f:R3_500_pd}, and \ref{f:pd_corot1k_all}, 
with the calculated 
depth offsets versus the input depths for synthetic eclipses put into the light curve using Star 2 from the CoRoT-1 Ks-band field. 
The green and pink points with solid lines represent the offsets with linear correlations only (same as the second plot in 
Figure~\ref{f:pd_corot1k_all}, but using the eclipse model of CoRoT-1b instead of Exoplanet X), while the 
blue and orange points with dashed lines show the results using quadratic correlation models for the positional displacement.

In these tests we found that the quadratic models underestimated the eclipse depths for all input depths, meaning that they were 
both more complex and less accurate. However, that may be the case only for this specific data set, and knowing that using 
a quadratic or more complicated fit model in place of a linear correlation might affect the depth estimation, it is recommended to 
at least perform this test on some of the field stars.

\section{Identifying and Measuring Real Eclipses}\label{sec:corot1}

Having tested the MCMC analysis described in the previous sections on a wide range 
of test eclipse light curves with both real and synthetic noise patterns, we then applied the 
process to re-evaluate the CoRoT-1b Ks-band eclipse depth measurement in R09. 
As we saw in Section~\ref{sec:realdata}, 
each light curve had  
its own systematics that caused a different bias for each star when treated as the target. 
We measured the eclipse signals in the target light curve as well as other field stars 
in multiple ways, in order to establish a corrected, more robust value for the 
true eclipse depth.

\subsection{Measurement of the CoRoT-1b Eclipse}\label{ss:corot1}

The main analysis of R09 optimized the differential photometry between the target star CoRoT-1 
and four nearby reference stars (those numbered 1, 2, 3, and 4 in this analysis), 
and then tested the fitting and removal of trends manually, one at a time. 
That analysis found that correcting for a correlation between the out-of-eclipse 
differential flux and the FWHM of the star images was sufficient to reach a 
detection of an eclipse of depth 
$D_e=3360\pm$420 ppm, centered at phase $\phi_{me}=0.5022^{+0.0023}_{-0.0027}$.

 For consistency in re-measuring the eclipse, we took the weighted differential light curve 
 with respect to the same four comparison stars 
 (without adding any eclipse signal). 
  The resulting light curve had a flux 
 RMS of 5390 ppm, and measured white and red noise levels 
 (fitting to Equation \ref{e:rednoise}) of $\sigma_w$=7320 ppm, $\sigma_r$=1230 ppm. 
 Looking only at the out-of-eclipse baseline points, however, the noise levels drop, with the 
 red noise component nearly vanishing: 
$\sigma_w$=6610 ppm, $\sigma_r$=170 ppm.

We then ran the MCMC-A analysis process to find the best-fit model incorporating a secondary 
eclipse signal -- now using 
the previously-measured characteristics for the planet CoRoT-1b by its host star 
(see Table~\ref{t:exopx}) rather than ``Exoplanet X'' -- and allowing correlations with any of the other variables 
in Table~\ref{t:mplist}. 
The best-fit eclipse model using the $\chi^2$ criterion included an eclipse 
$3240^{+330}_{-480}$ ppm deep, centered at phase $0.50410^{+0.00077}_{-0.00072}$; with the 
BIC, the best-fit model had $D_e=3150^{+380}_{-400}$ ppm, $\phi_{me}=0.50426^{+0.00062}_{-0.00094}$.
These results, all agreeing with the previously published values within the reported errors of R09, 
are listed in Table~\ref{t:realresults} in the column marked ``0-a''.

The small uncertainties on $\phi_{me}$ appear to 
suggest an eccentricity that is nonzero at the 4 to 5-$\sigma$ level; 
however, the phase values for the real datasets were calculated based 
on the ephemeris from \citet{Bean2009}, derived from multiple transit timing observations: 
$T_0$(RJD)$=54159.452879\pm0.000068$, $p=1.5089656\pm0.000006$ d. 
Propogating these errors and adding them into our uncertainties,
we adjust the best-fit values of $\phi_{me}$ to 
0.50410$^{+0.00197}_{-0.00195}$ ($\chi^2$) and 0.50426$^{+0.00191}_{-0.00204}$ (BIC). 
Both of these now show a nonzero eccentricity 
at only a 2.1-$\sigma$ level, 
and place the R09 value of $\phi_{me}$=0.5022 at 1~$\sigma$ 
away from the new measurements.

Because the two information criteria methods gave nearly identical results, we determined it best to adopt 
average values of the two solutions, with the error bars adjusted to include the spread of both. Thus, we state 
as our re-measurement of the R09 eclipse the values $D_e$=3190$^{+370}_{-440}$ and 
$\phi_{me}$=0.50418$^{+0.00197}_{-0.00203}$.

We also tried constructing other light curves using different combinations of stars 1 through 4 as the comparison 
flux for the target CoRoT-1. The measurements resulting from one comparison star were inconsistent, sometimes 
detecting a negative depth or strongly offset phases, and high levels of red noise in the residuals. 
However, there was better consistency in the tests that used two or more reference stars, with a most of the 
best-fit eclipse models having a depth between 2800 and 3500 ppm and a central phase of 0.502 to 0.504.

Although these measurements produced eclipse attributes consistent with the results of R09, 
we saw in the analysis for Section~\ref{ss:rdresults} that it was common to find eclipse solutions of significantly 
non-zero depth in the light curves of other field stars with no synthetic eclipse added. 
Two examples of this are seen in the best-fit results for light curves 1-a and 4-a, run through the same process 
as the target (the third and fourth columns of Table~\ref{t:realresults}), and examining the results from 
test target stars 1, 2, 3, 4, 8, and 13, we noted two important findings.
First, the detected depths 
appeared to be spread in a non-Gaussian distribution; 
rather than finding many of the depths close to zero with few outliers, they were clustered close to the red noise amplitude 
in both the positive and negative direction. 
Second, most of the spurious detections were not centered at phases close to 0.5 -- only one of the 
positive eclipse values (4-a with BIC) had a $\phi_{me}$ within even 5~$\sigma$ of 0.5. 
These spurious detections make it difficult to determine how much a detected eclipse signal in the target light curve 
is the result of systematic effects versus an actual measurement of the planet flux.
In particular, since the reliability of the measurements is poorest for eclipse signals below 
the noise levels, eclipses of small depth are very difficult to measure.

\subsection{Comparison to an Independent Reference Set}\label{ss:groupb}

In all of the real noise tests, the systematics were produced by a combination of the target star's light curve and the 
composite flux of the different reference stars, so a bias could result from either the target or the reference 
star flux.
To create a more controlled dataset to compare the detection from CoRoT-1 with 
the spurious detections from the other test stars, we prepared an external ``Group B'' of stars 
(numbers 5, 6, 7, 8, 9, 11, 12, and 13 in the field), and combined their fluxes into a single composite light curve that we could 
use to test each star from ``Group A'' (stars 1, 2, 3, 4, and CoRoT-1) in a consistent manner. With these we created five new 
differential light curves, ``0-b'' through ``4-b''. Since we were not using the optimal set of reference stars, their noise 
levels were approximately 1.5 to 2 times higher than for the ``A'' set of light curves.

We then ran these light curves through the MCMC-A process as before; the right-hand side of Table~\ref{t:realresults} lists 
three of them -- 0-b, 1-b, and 4-b --  with their calculated white and red noise amplitudes (in ppm) and the results fo their 
best-fit solutions. We saw similar behavior to what appeared in the Group A light curve results: 
eclipse depths at large values, and mid-eclipse phases often far from 0.5. 
The depths of these spurious eclipse detections were more consistent in these tests than in the Group A tests, 
clustered mainly at 4100 to 4800 ppm, and with only one negative depth detection among them. 
The measurement of the eclipse on the target light curve 0-b did not plainly stand out from the rest; its 
depth using $\chi^2$ was only slightly greater than that for 4-b, and while its $\phi_{me}$ values were the 
closest to 0.5, they were not much closer than the values found for 1-b.

The eclipse depths found at large phase offsets, however, tell us little about the bias that 
the comparison set may contribute when combined with a real eclipse. 
When a planet's orbital eccentricity is known from other measurements, as is the case 
with CoRoT-1b, that information can be used to better rule out false eclipse detections. 
We expected the true eclipse to be centered close to $\phi$=0.5 
(recall R09 measured $0.5022^{+0.0023}_{-0.0027}$, and the detections in other bands either 
had similar results or assumed a circular orbit), and thus tried running additional MCMCs with that parameter fixed. 
Table~\ref{t:realr2} lists these results for fixing $\phi_{me}$ to both 0.5 and 0.5022, and 
in each of these cases, we saw new trends. 
Since the routine could no longer fit deep eclipses at phases far from 0.5, 
the eclipses measured in the light curves from stars 1 through 4 were shallower than before. 
There also were no negative depths measured, and the eclipse signal for light curve 3-b was nearly zero.

The greatest depth measurements were now by a significant margin 
those for the target light curve 0-b, ranging from 
4310$^{+700}_{-410}$ to 5780$^{+380}_{-580}$ ppm. 
These values are significantly higher than in R09; however, there was a
greater amount of noise in the comparison flux (composite of the eight ``Group B'' stars), 
than in the optimal reference composite from stars 1 through 4 (``Group A''), and the uniformly positive 
spurious detections in 1-b through 4-b suggest that some of this depth is a bias in the comparison 
light curve. 
Conveniently, we can look at these results for the other light curves with the same 
comparisons to determine what bias they may be introducing.

For each fixed $\phi_{me}$ and information criterion, we estimated the Group B bias 
contribution with a weighted average of the eclipse depths found for target stars 1 through 4. The 
weighting factors were the same that were used for the composite comparison flux from Group A: 
40\% for 2-b, 22\% for 3-b, and 19\% each for 1-b and 4-b. Then to determine a corrected depth 
measurement for the real eclipse, we simply subtracted the bias measurement from the best-fit 
0-b light curve result. The final, corrected $D_e$ values for each fixed-phase test are given (in ppm) 
in the last row of Table~\ref{t:realr2}. 
These values end up in fairly close agreement with one 
another as well as with the R09 value. In particular, the best-fit depths 
using $\chi^2$ are nearly identical 
(3210$^{+560}_{-590}$ ppm at $\phi_{me}$=0.5, 3240$^{+530}_{-520}$ ppm at $\phi_{me}$=0.5022), 
suggesting that the measured flux decrement is not 
strongly dependent on the exact placement of the mid-eclipse phase in these trials. 
The BIC results are less consistent with one another 
(2650$^{+760}_{-500}$ ppm at $\phi_{me}$=0.5, 3750$\pm620$ ppm at $\phi_{me}$=0.5022), but 
remain in agreement within their errors of the $\chi^2$ results, the values in Section~\ref{ss:corot1}, 
and the original measurement in R09.

These additional test results served to reinforce that the measurements of CoRoT-1b are the most consistent in our 
photometric dataset; the eclipse was found at close to the same depth and phase placement by a 
number of different means. 
The red noise was eliminated or nearly eliminated in many of the residuals, indicating that the process did 
a good job of removing trends.


\section{Summary \& Discussion}\label{sec:discussion}

Much of our understanding of exoplanet atmospheres is based on a small number of 
emission detections, making it crucial to know how accurate those observations are.  
Especially when using ground-based instruments, correlated  ``red'' noise complicates obtaining
a reliable measurement.  
Measurements subject to red noise require an optimization of 
the observations to limit the systematic effects, and then the identification, modeling, and removal
of the trends from the resulting light curves.  
Simultaneously considering the eclipse shape and correlation trends together in a single model 
often requires a large number of parameters to be optimized, making use of statistical 
tools like the Bayesian Information Criterion (BIC) and a Markov Chain Monte Carlo (MCMC) algorithm.

To test the reliability of commonly used analysis methods, 
we created a sample of hundreds of test eclipse light curves with both synthetic and real noise 
and ran them through the processes we developed for modeling (the combination of the eclipse signal 
and trends) and parameter optimization (MCMC). 
To simulate red noise, we used both simple sinusoidal patterns and stochastic Brownian noise, 
observing similar results with each. 
Another concern regarding the synthetic models was that the noise amplitude was time-invariant, 
whereas in actual data sets the changes in airmass or sky brightness, or instrumental changes like 
defocusing will change the noise level over the course of the observation. This assumption 
of stationarity is used in all of the current modeling approaches, but nevertheless may lead 
to an underestimation of the effects of real noise. 
A selection of these data sets are made available in Appendix~\ref{sec:benchmark}
 for testing other analysis methods.
From our tests and results, we determined some lessons for both the 
observation and analysis process.

In the case where only white noise is present in the light curves, 
our tests find for each sampling rate, baseline, and information criterion a critical input depth 
dividing the results. 
Eclipse signals with input depths greater than this value are generally recovered within 10\% without any
systematic biases, while below it the eclipse depth measurements differ from the input depths 
by more than their measured uncertainties. This critical depth can be found by beginning with 
synthetic eclipses of a great depth relative to the white noise level, and testing lesser depths 
until they are no longer measured within the desired accuracy. 
We find that the sampling rate and baseline are both important considerations: a longer baseline 
and a faster cadence improve the limits to which one can make a successful measurement and reduce the dispersion of the errors. 

When red noise is present in the light curve, the dispersion of results is greater and the critical depth is reached at 
a higher input value; more importantly, we find that several of the series exhibit systematic biases. 
Whereas in the white noise tests, the discrepancy between the input and measured eclipse depths and 
mid-eclipse phases was randomly positive or negative, 
structure in the red noise often results in measured eclipses that are consistently 
deeper or shallower than the inputs, or centered at a particular phase offset. 
The uncertainties reported by the MCMC also may underestimate the depth errors by a factor of two to four.
We find that baselines that significantly exceed the duration of the eclipse are still usually preferable; 
however, in certain cases the structure of the noise can negatively affect the best-fit solutions. 
Including all available baseline may not be optimal for real datasets either, particularly if 
some of it is noisier (e.g.~taken at high airmass or different weather conditions). 
In cases when this is possible, additional tests should be done with a clipped baseline.

Examining datasets with real noise, we find unexplained or under-modeled systematics to be more prevalent still, 
which can lead to large biases or even completely spurious detections.
Even with a detailed model to account for a wide variety of trends in the light curves, 
we find that discrepant results between input and output signals appear when using different field stars for photometric comparison.
For this reason, confirming a measurement using different sets of comparison stars is advised. 
Much larger biases can accumulate on shallow eclipses than on deeper ones, and 
each star shows its own systematics and biases that sometimes cannot be removed by de-correlation techniques.
The actual depth biases can be anywhere from about the same magnitude as the measured uncertainties to 
three to four times as large, reinforcing the common attitude of preferring substantially more than a 3-$\sigma$ detection.
A significant finding is that spurious eclipse detections measured on field stars do not follow a 
Gaussian distribution, but rather cluster around the level of the light curve's red noise, with depths in 
both the positive and negative direction.

This uncertainty of what is a real eclipse signal and what is a false detection from red noise 
poses a serious problem to the measurement of eclipses, and 
further tests should be used to determine the nature of the biases and verify the detections.
One way to do so is to design a control set of comparison stars and determine a bias level by combining 
the spurious detections of other stars in place of the target. If the expected mid-eclipse time (i.e.~eccentricity) 
is well-known, this value can be fixed to better determine the noise structure's contribution to the 
measurement of the actual eclipse depth. The established bias level can then be corrected for 
in the eclipse measurement of the actual target to verify its depth.

These additional tests and verification processes are particularly advisable in the case of low-confidence 
($<10\sigma$) detections, i.e.~most of the ground-based measurements that have been published.
In order to claim a confident detection, 
we recommend the following criteria be met:
\begin{itemize}
\item A sufficient level of significance on the measured signal (commonly adopted as $\sim 5\sigma$), 
and agreement using different comparison stars and/or methods
\item Mid-eclipse phase consistent with other studies of the eccentricity
\item A relatively small amount of red noise in the residuals
\item Spurious eclipse signals with similar confidence levels and phase placements as the target not seen 
in the light curves of other nearby stars
\end{itemize}
This set of criteria is stricter than adopted in the literature, but ensures more reliable detections and 
depth measurements. 
We provide as an example the tests performed on the CoRoT-1 data from R09.
Whereas the tests on other stars in the field produce a distribution 
of eclipse detections with positive and negative depths and a range of mid-eclipse phases suggesting 
strongly nonzero eccentricities, the measurements of the eclipse in the CoRoT-1 light curve agree 
closely in depth and with a mid-eclipse phase close to 0.5.
Testing different light curves for the target using different reference stars, the results are not 
strongly affected as long as at least two of the bright, nearby comparison stars are used. 
When using a separate set of less-ideal reference stars, a deeper eclipse is measured, but this bias is 
estimated and removed effectively by finding the depth measured in other stars with the same reference 
set, fixed to the same mid-eclipse phase.
Through this variety of tests, we establish the R09 detection to be robust and consistent.

Given the difficulties and complications of measuring secondary eclipse signals in typical 
ground-based data, we caution that both published and working detections be checked to ensure 
that they represent an actual eclipse measurement and are not the products of unresolved biases. 
If the published eclipse depths (and, subsequently, planetary thermal emission levels) are adjusted, 
the best-fitting models and conclusions about atmospheric composition and structure can be dramatically altered. 
We recommend that these analysis routines and tests be applied to current and future eclipse light curves, 
as a means to evaluate and minimize the presence of systematics in the reported eclipse signals.

\acknowledgments{This research has been supported by the National Science 
Foundation through grant 
AST-0908278. Critical support was provided by the Space Telescope Science Institute 
Director's Discretionary Research Fund D0101.90131. We thank the scientific editor and the anonymous 
referee for insightful comments and suggestions, and Veselin Kostov 
for helpful discussion.}


\bibliographystyle{apj}
\bibliography{Roge0202}

\clearpage




\begin{table}[]
\begin{center}
\caption{Characteristics of Exoplanet X and CoRoT-1b. 
{\bf References.} 
(1) See Section~\ref{ss:lcconstruction}
(2) \citealt{Barge2008}
(3) \citealt{Bean2009}
\label{t:exopx}
}
\begin{tabular}{l|cccc}
\hline
Parameter & Exoplanet X & CoRoT--1b \\
\hline
Star Radius $R_*$ [$R_{Sun}$] & 1.0 & 0.95 \\
Planet Radius $R_P$ [$R_{Jup}$] & 1.5 & 1.49 \\
Period $P$ [days] & 1.041667 & 1.508956 \\
Semimajor Axis $a$ [AU] & 0.01978 & 0.02540 \\
Inclination $i$ [degrees] & 88.0 & 85.1 \\
Eccentricity $e$ & 0 & 0\\
$T_{14}$ [minutes] & 130.15 & 150.32 \\
$T_{23}$ [minutes] & 94.13 & 106.56 \\
Reference & 1 & 2, 3 \\
\hline
\end{tabular}
\end{center}
\end{table}

\begin{table}[]
\begin{center}
\caption{Parameters Used for Light Curve Models
\label{t:mplist}
}
\begin{tabular}{lc}
\hline
Model Parameter & Symbol\\
\hline
Eclipse Depth & $D_e$\\
Mid-Eclipse Phase & $\phi_{me}$\\
Baseline Flux Level & $F_b$\\
Phase Slope & $m_{\phi}$\\
Airmass Slope & $m_z$\\
FWHM Slope & $m_w$\\
X-displacement Slope & $m_x$\\
Y-displacement Slope & $m_y$\\
Sky Brightness Slope & $m_s$\\
\hline
\end{tabular}
\end{center}
\end{table}

\begin{sidewaystable}[]
\begin{center}
\caption{Results from Two Independent MCMC Methods 
\label{t:jrvsea}
}
\begin{tabular}{ccc|ccc|ccc}
\hline
$\sigma_r$ (ppm) & Obs. Win & Samp. Rate & \multicolumn{3}{|c|}{Eclipse Depth $D_e$ (ppm)} & 
\multicolumn{3}{|c}{Mid-eclipse phase $\phi_{me}$} \\
(pattern) & (minutes) & (points/hour) & Input & Fit (A) & Fit (B) & Input & Fit (A) & Fit (B) \\
\hline
0 & 240 & 25 & 1300 & 1318$^{+141}_{-180}$ & 1210$\pm$210 & 
  0.500 & 0.50093$^{+0.00181}_{-0.00132}$ & 0.50094$\pm$0.00257\\
0 & 506 & 75 & 1600 & 1502$^{+79}_{-59}$ & 1470$\pm$99 & 
  0.500 & 0.50123$^{+0.00072}_{-0.00111}$ & 0.50035$\pm$0.00093\\ 
0 & 240 & 250 & 1600 & 1599$^{+31}_{-62}$ & 1520$\pm$65 & 
  0.500 & 0.49914$^{+0.00046}_{-0.00031}$ & 0.49934$\pm$0.00048\\
200 (P1) & 240 & 75 & 2000 & 2261$^{+130}_{-46}$ & 2200$\pm$119 & 
  0.500 & 0.50176$^{+0.00055}_{-0.00055}$ & 0.50286$\pm$0.00050\\
500 (P2) & 240 & 75 & 2000 & 1757$^{+88}_{-87}$ & 1670$\pm$120 & 
  0.500 & 0.49940$^{+0.00093}_{-0.00062}$ & 0.49887$\pm$0.00060\\
1000 (P3) & 240 & 75 & 2000 & 2367$^{+72}_{-98}$ & 2320$\pm$119 & 
  0.500 & 0.50207$^{+0.00053}_{-0.00061}$ & 0.50257$\pm$0.00063\\
0 & 240 & 75 & 2300 & 2128$^{+100}_{-70}$ & 2040$\pm$120 & 
  0.500 & 0.49929$^{+0.00067}_{-0.00046}$ & 0.49930$\pm$0.00162\\
200 (P3) & 240 & 250 & 4000 & 4014$^{+51}_{-43}$ & 3890$\pm$66 & 
  0.500 & 0.49994$^{+0.00019}_{-0.00013}$ & 0.49982$\pm$0.00024\\
\hline
\end{tabular}
\end{center}
\end{sidewaystable}

\begin{table}[]
\begin{center}
\caption{Synthetic Light Curve Parameters Tested
\label{t:synthparams}
}
\begin{tabular}{l|l}
\hline
Parameter & Values Tested\\
\hline
White Noise Level $\sigma_w$ (ppm) & 1000\\
Red Noise Pattern (See Fig.~\ref{f:rnmodels}) & 0, 1, 2, 3\\
Red Noise Level $\sigma_r$ (ppm) & 200, 500, 1000\\
Observing Window (minutes) & 160, 240, 318, 506 \\
Input Mid-Eclipse Phase $\phi_{me}$ & .475, .500\\
Sampling Rate (points / hour) & 25, 75, 250\\
Input Eclipse Depth $D_e$ (ppm) & 50, 100, 150, 200, 250, 300, 400, 500, \\ 
& 600, 800, 1000, 1300, 1600, 2000, 2300, \\ 
& 2600, 3000, 3300, 3600, 4000, 4500, 5000\\
Information Criterion & $\chi^2$, BIC\\
\hline
\end{tabular}
\end{center}
\end{table}


\begin{sidewaystable}[t]
\begin{center}
\caption{Example Results of MCMC Trials
\label{t:exresults15}
}
\begin{tabular}{c | cc | ccc | cc }
\hline
 & \multicolumn{2}{|c|}{MCMC Starting Point} & \multicolumn{3}{|c|}{Recovered Values} & & Residuals\\
Trial & $D_e$ (ppm) & $\phi_{me}$ & $D_e$ (ppm) & $\phi_{me}$ & $F_b$ & $\chi^2$ & RMS (ppm)\\
\hline
1-1 & \multirow{3}{*}{        -744} & \multirow{3}{*}{  0.50311} &  $        4390^{+          24}_{         -76}$
 & $  0.50080^{+  0.00015}_{ -0.00022}$ & $  1.000589^{+  0.000053}_{ -0.000019}$ &   758.032 &         1068\\
1-2 & & & $        4381^{+          31}_{         -68}$ & $  0.50078^{+  0.00017}_{ -0.00020}$
 & $  1.000593^{+  0.000049}_{ -0.000022}$ &   757.966 &         1068\\
1-3 & & & $        4365^{+          47}_{         -52}$ & $  0.50073^{+  0.00022}_{ -0.00015}$
 & $  1.000626^{+  0.000017}_{ -0.000056}$ &   758.216 &         1068\\
\hline
2-1 & \multirow{3}{*}{        6724} & \multirow{3}{*}{  0.51383} & $        4348^{+          65}_{         -36}$
 & $  0.50081^{+  0.00015}_{ -0.00022}$ & $  1.000594^{+  0.000049}_{ -0.000024}$ &   758.347 &         1068\\
2-2 & & & $        4354^{+          58}_{         -42}$ & $  0.50070^{+  0.00025}_{ -0.00012}$
 & $  1.000611^{+  0.000032}_{ -0.000040}$ &   758.057 &         1068\\
2-3 & & & $        4388^{+          24}_{         -75}$ & $  0.50083^{+  0.00013}_{ -0.00024}$
 & $  1.000603^{+  0.000039}_{ -0.000033}$ &   758.190 &         1068\\
\hline
3-1 & \multirow{3}{*}{        4039} & \multirow{3}{*}{  0.47496}  & $        4360^{+          53}_{         -47}$
 & $  0.50071^{+  0.00025}_{ -0.00012}$ & $  1.000589^{+  0.000054}_{ -0.000019}$ &   758.262 &         1068\\
3-2 & & & $        4378^{+          35}_{         -66}$ & $  0.50073^{+  0.00023}_{ -0.00014}$ 
& $  1.000614^{+  0.000029}_{ -0.000044}$ &   758.185 &         1068\\
3-3 & & & $        4382^{+          30}_{         -69}$ & $  0.50084^{+  0.00012}_{ -0.00025}$
 & $  1.000613^{+  0.000030}_{ -0.000042}$ &   758.203 &         1068\\
\hline
4-1 & \multirow{3}{*}{       -4214} & \multirow{3}{*}{  0.47870}  & $       -4399^{+          98}_{        -102}$
 & $  0.45004^{+  0.00000}_{ -0.00004}$ & $  1.005478^{+  0.000095}_{ -0.000034}$ &  2915.416 &         2094\\
4-2 & & & $       -4456^{+         154}_{         -45}$ & $  0.45001^{+  0.00001}_{ -0.00001}$
 & $  1.005494^{+  0.000078}_{ -0.000048}$ &  2913.833 &         2094\\
4-3 & & &  $       -4414^{+         113}_{         -86}$ & $  0.45001^{+  0.00001}_{ -0.00001}$
 & $  1.005521^{+  0.000050}_{ -0.000076}$ &  2912.509 &         2093\\
\hline
5-1 & \multirow{3}{*}{       -4394} & \multirow{3}{*}{  0.51646}  & $        4362^{+          51}_{         -49}$
 & $  0.50084^{+  0.00011}_{ -0.00026}$ & $  1.000610^{+  0.000033}_{ -0.000039}$ &   757.959 &         1068\\
5-2 & & &  $        4391^{+          22}_{         -77}$ & $  0.50082^{+  0.00013}_{ -0.00024}$
 & $  1.000609^{+  0.000034}_{ -0.000038}$ &   758.231 &         1068\\
5-3 & & & $        4388^{+          25}_{         -75}$ & $  0.50073^{+  0.00022}_{ -0.00015}$
 & $  1.000621^{+  0.000022}_{ -0.000051}$ &   758.585 &         1068\\
\hline
5-1 & \multicolumn{2}{|c|}{(Best Trial, Adjusted)} & $        4359^{+          51}_{         -49}$
 & $  0.50084^{+  0.00011}_{ -0.00026}$ & 1.000000 &   757.959 &         1068\\
\hline
\end{tabular}
\end{center}
\end{sidewaystable}

\begin{sidewaystable}[]
\begin{center}
\caption{Synthetic Series Critical Depths and Dispersions
\label{t:cdd}
}
\begin{tabular}{ccc|cc|cc|cc|cc}
\hline
Obs. & Sampling & & \multicolumn{2}{c|}{$\sigma_r$=0} & \multicolumn{2}{c|}{$\sigma_r$=200 ppm} &
\multicolumn{2}{c|}{$\sigma_r$=500 ppm} & \multicolumn{2}{c}{$\sigma_r$=1000 ppm}\\
Window & Rate & Info. & $D_{e,crit}$ & Disp. & $D_{e,crit}$ & Disp. & 
$D_{e,crit}$ & Disp. & $D_{e,crit}$ & Disp.\\
(minutes) & (pts./hr.) & Crit. & (ppm) & (\%) & (ppm) & (\%) & (ppm) & (\%) & (ppm) & (\%)\\
\hline
160 & 25 & $\chi^2$ & 3600 & 7.00 & 4000 & 2.41 & 3000 & 6.49 & $>5000$ & 23.66\\
160 & 75 & $\chi^2$ & 1000 & 3.95 & 3000 & 3.47 & 3600 & 8.35 & $>5000$ & 19.17\\
160 & 250 & $\chi^2$ &1000 & 3.71 & 2300 & 4.31 & 4000 & 9.35 & $>5000$ & 13.73\\
\hline
160 & 25 & BIC & 2000 & 4.25 & 3600 & 2.78 & 4000 & 8.74 & 4000 & 6.34\\
160 & 75 & BIC & 1000 & 3.90 & 2300 & 3.05 & 3600 & 6.36 & $>5000$ & 11.05\\
160 & 250 & BIC& 400 & 3.71 & 2000 & 6.62 & 3600 & 8.14 & $>5000$ & 11.63\\
\hline
240 & 25 & $\chi^2$ & 800 & 3.84 & 2000 & 5.59 & 2300 & 5.56 & 3000 & 5.94\\
240 & 75 & $\chi^2$ & 1000 & 2.29 & 1000 & 4.16 & 2000 & 2.96 & 4000 & 5.22\\
240 & 250 & $\chi^2$ & 800 & 2.41 & 600 & 4.01 & 2000 & 4.14 & 2300 & 8.59\\
\hline
240 & 25 & BIC & 800 & 5.98 & 3300 & 4.78 & 1300 & 5.90 & $>4500$ & 6.58\\
240 & 75 & BIC & 1000 & 3.08 & 1000 & 4.80 & 2300 & 5.83 & 3300 & 5.75\\
240 & 250 & BIC & 800 & 3.33 & 800 & 3.47 & 1600 & 5.54 & 3000 & 7.68\\
\hline
318 & 25 & $\chi^2$ & 2000 & 3.90 & 2300 & 2.21 & 1300 & 4.43 & 800 & 5.66\\
318 & 75 & $\chi^2$ & 500 & 3.41 & 1300 & 1.96 & 500 & 3.38 & 1300 & 3.65\\
318 & 250 & $\chi^2$ & 500 & 2.91 & 400 & 1.59 & 250 & 2.86 & 800 & 2.86\\
\hline
318 & 25 & BIC & 1600 & 3.20 & 600 & 4.37 & 2600 & 5.78 & 1600 & 4.75\\
318 & 75 & BIC & 800 & 4.62 & 500 & 3.75 & 800 & 4.37 & 600 & 3.44\\
318 & 250 & BIC & 250 & 3.59 & 800 & 2.99 & 800 & 2.65 & 1000 & 2.92\\
\hline
506 & 25 & $\chi^2$ & 800 & 4.82 & 800 & 4.72 & 2300 & 6.29 & $>5000$ & 13.82\\
506 & 75 & $\chi^2$ & 300 & 3.96 & 1300 & 4.83 & 3600 & 8.98 & $>5000$ & 14.25\\
506 & 250 & $\chi^2$ & 300 & 2.80 &1600 & 4.75 & 3000 & 7.66 & $>5000$ & 13.22\\
\hline
506 & 25 & BIC & 1600 & 3.54 & 2300 & 5.42 & 3000 & 8.38 & $>5000$ & 15.57\\
506 & 75 & BIC & 300 & 3.56 & 2000 & 4.24 & 2600 & 6.45 & $>4500$ & 10.96\\
506 & 250 & BIC & 300 & 2.17 & 1000 & 5.10 & 3000 & 8.61 & $>5000$ & 14.09\\
\hline
\end{tabular}
\end{center}
\end{sidewaystable}

\begin{sidewaystable}[]
\begin{center}
\caption{Real Noise Data Sets and Results
\label{t:realresults}
}
\begin{tabular}{l|ccc|cccc}
\hline
Light Curve & 0-a & 1-a & 4-a & 0-b & 1-b & 4-b\\
\hline
Target & CoRoT-1 & Star 1 & Star 4 & CoRoT-1 & Star 1 & Star 4\\
Comp. Stars & 1,2,3,4 & 2,3,4,13 & 1,2,3,8,13 & Group B & Group B & Group B\\
$\sigma_w$ (all) & 7318 & 4582 & 4550 & 8265 & 8982 & 9283\\
$\sigma_r$ (all) & 1225 & 1535 & 1691 & 2257 & 1451 & 5437\\
$\sigma_w$ (out) & 6608 & 4462 & 4378 & 7446 & 8451 & 9329\\
$\sigma_r$ (out) & 171 & 1609 & 1903 & 1527 & 1878 & 3688\\
\hline
 & \multicolumn{6}{c}{\emph{Best-fit Model Results ($\chi^2$)}} & \\
\hline
$D_e$ (ppm) & $3237^{+327}_{-477}$ &  $-3345^{+513}_{-604}$ & $2769^{+495}_{-862}$ & 
$5798^{+375}_{-406}$ & $4634^{+436}_{-386}$ & $5600^{+542}_{-629}$ \\
$\phi_{me}$ & $0.50410^{+0.00077}_{-0.00072}$ & $0.45440^{+0.00088}_{-0.00153}$ & $0.52223^{+0.00106}_{-0.00307}$ & 
$0.50468^{+0.00049}_{-0.00047}$ & $0.50681^{+0.00174}_{-0.00078}$ & $0.52041^{+0.00061}_{-0.00073}$ \\
Nonzero params & 15 & 16 & 16 & 16 & 16 & 16\\
$\sigma_w$ (residuals) & 6544 & 5660 & 5505 & 7177 & 7951 & 7773\\
$\sigma_r$ (residuals) & 125 & 0 & 606 & 0 & 0 & 646\\ 
\hline
 & \multicolumn{6}{c}{\emph{Best-fit Model Results (BIC)}} & \\ 
\hline
$D_e$ (ppm) & $3150^{+383}_{-401}$ & $-3492^{+829}_{-414}$ & $1559^{+354}_{-1463}$ & 
$6007^{+434}_{-406}$ & $4188^{+311}_{-288}$ & $4289^{+457}_{-353}$\\
$\phi_{me}$ & $0.50426^{+0.00062}_{-0.00094}$ & $0.45449^{+0.00101}_{-0.00157}$ & $0.50009^{+0.00183}_{-0.00515}$ & 
$0.50458^{+0.00052}_{-0.00042}$ & $0.50721^{+0.00131}_{-0.00080}$ & $0.52018^{+0.00080}_{-0.00074}$\\
Nonzero params & 15 & 10 & 9 & 12 & 8 & 8\\
$\sigma_w$ (residuals) & 6411 & 5617 & 5498 & 7280 & 8056 & 7944\\
$\sigma_r$ (residuals) & 0 & 0 & 781 & 0 & 335 & 522\\
\end{tabular}
\end{center}
\end{sidewaystable}

\begin{sidewaystable}[]
\begin{center}
\caption{Independent Reference Set (Group B) Tests and Corrections
\label{t:realr2}
}
\begin{tabular}{c|cc|ccc}
\hline
 & \multicolumn{2}{c|}{\emph{$\phi_{me}$ fixed to 0.5000}}
 & \multicolumn{2}{c}{\emph{$\phi_{me}$ fixed to 0.5022}}\\
Target & Best-Fit $D_e$ ($\chi^2$) & Best-Fit $D_e$ (BIC)
 & Best-Fit $D_e$ ($\chi^2$) & Best-Fit $D_e$ (BIC)\\
\hline
CoRoT-1 & $4992^{+430}_{-403}$ & $4312^{+699}_{-405}$  
 & $5556^{+426}_{-365}$   & $5784^{+375}_{-584}$  \\      
\hline
Star 1 & $3861^{+332}_{-525}$ & $3691^{+280}_{-350}$
 & $4103^{+452}_{-348}$    & $3837^{+306}_{-312}$   \\
Star 2 & $1618^{+268}_{-347}$    & $1575^{+340}_{-293}$
 & $2389^{+290}_{-332}$   & $2342^{+696}_{-181}$   \\
Star 3 & $101^{+462}_{-479}$   & $0^{+36}_{-36}$
 & $634^{+427}_{-441}$   & $46^{+24}_{-101}$   \\
Star 4 & $2003^{+427}_{-436}$   & $1721^{+322}_{-358}$
 & $2312^{+460}_{-396}$    & $1869^{+527}_{-211}$   \\
\hline
Avg. 1-4 & $1784^{+362}_{-433}$ & $1658^{+285}_{-287}$ 
 & $2314^{+391}_{-374}$ & $2031^{+514}_{-206}$ \\
\hline
Corrected & $3208^{+562}_{-591}$ & $2654^{+755}_{-496}$  
 & $3242^{+528}_{-522}$  & $3753^{+622}_{-619}$\\
\hline
\end{tabular}
\end{center}
\end{sidewaystable}

\begin{table}[]
\begin{center}
\caption{Benchmark Data Set Attributes
\label{t:benchmarks}
}
\begin{tabular}{cccccccc}
\hline
Name & Type & $\sigma_w$ & $\sigma_r$ & Obs. Win & Samp. Rate & $D_e$ & $\phi_{me}$\\
\hline
S1 & Synth & 1000 & 0 & 240 & 75 & 4000 & 0.500\\
S2 & Synth & 1000 & 0 & 240 & 75 & 3300 & 0.475\\
S3 & Synth & 1000 & 0 & 160 & 75 & 1000 & 0.500\\
S4 & Synth & 1000 & 0 & 506 & 75 & 1600 & 0.500\\
S5 & Synth & 1000 & 200 & 240 & 75 & 2000 & 0.500\\
S6 & Synth & 1000 & 1000 & 240 & 250 & 4000 & 0.500\\
S7 & Synth & 1000 & 200 & 318 & 25 & 2600 & 0.500\\
R1 & Star 2 & 6188 & 2239 & 293 & 141 & 4000 & 0.500\\
R2 & Star 0 & 7318 & 1225 & 293 & 141 & 1500 & 0.500\\
R3 & Star 0 & 7318 & 1225 & 293 & 141 & 0 & 0.500\\
\hline
\end{tabular}
\end{center}
\end{table}

\begin{table}[ht]
\begin{center}
\caption{Data Set S1: Synthetic Noise (Full Table Online)
\label{t:synthlc}
}
\begin{tabular}{ccc|cc|cc}
\hline
Phase &    O.Flux & Error & E.Shape &  N.Base & W.Noise & R.Noise\\
\hline
 0.42000   & 1.00452   & 0.00099   & 1.0040   & 1.00052   & 0.00052   & 0.00000 \\
 0.42053   & 1.00393   & 0.00099   & 1.0040   & 0.99993   & -0.00007   & 0.00000\\
 0.42107   & 1.00404   & 0.00099   & 1.0040   & 1.00004   & 0.00004   & 0.00000\\
 0.42161   & 1.00460   & 0.00099   & 1.0040   & 1.00060   & 0.00060   & 0.00000\\
 0.42214   & 1.00330   & 0.00099   & 1.0040   & 0.99930  & -0.00070   & 0.00000\\
 ... & ... & ... & ... & ... & ... & ...\\
\hline
\end{tabular}
\end{center}
\end{table}

\begin{sidewaystable}[ht]
\begin{center}
\caption{Data Set R1: Real Noise (Full Table Online)
\label{t:reallc}
}
\begin{tabular}{ccc|cc|c|ccccc}
\hline
Phase &    O.Flux & Error & E.Shape &  N.Base & O.Pos &  Airm.   &  FWHM     &   x Disp   &   y Disp   &   Sky \\
\hline
   0.43242 &      1.00245 &      0.00418 &      1.0040 &      0.99846 &         0 &         1.4050 &     10.6442 &      4.9667 &     -1.8051 &      -38.174\\
   0.43279 &      1.00067 &      0.00405 &      1.0040 &      0.99668 &         1 &         1.4023 &     10.6731 &      5.8475 &     -2.4504 &       35.485\\
   0.43294 &      1.00390 &      0.00411 &      1.0040 &      0.99990 &         1 &         1.4010 &     10.6792 &      5.8099 &     -2.4756 &       14.224\\
   0.43315 &      0.99956 &      0.00415 &      1.0040 &      0.99557 &         0 &         1.3995 &     10.6743 &      5.3403 &     -2.8960 &       -7.687\\
   0.43331 &      0.99974 &      0.00415 &      1.0040 &      0.99576 &         0 &         1.3983 &     10.6720 &      4.6221 &     -2.7726 &       -2.658\\
    ... & ... & ... & ... & ... & ... & ... & ... & ... & ... & ...\\
\hline
\end{tabular}
\end{center}
\end{sidewaystable}

\clearpage


\begin{figure}
{\centering
  \includegraphics[width=0.9 \textwidth,angle=0]{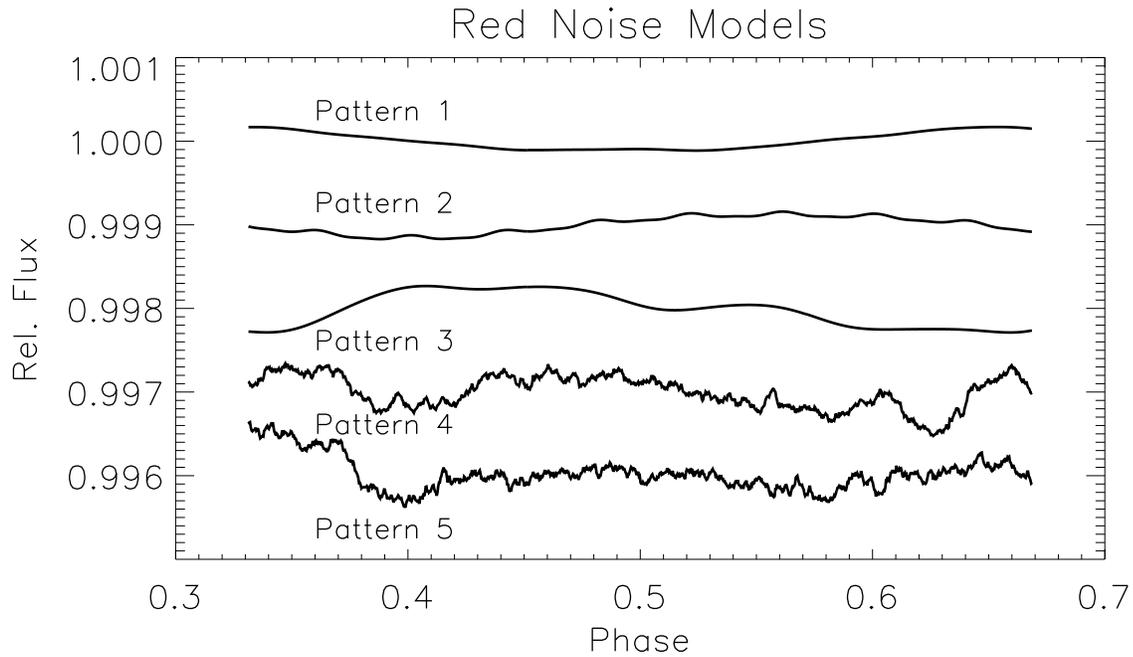}}
\caption[]{
Five red noise models we used for the synthetic light curve tests.
Patterns 1, 2, and 3 were constructed by adding together four 
sinusoidal waves of different periods, amplitudes, and phase angles.  
Patterns 4 and 5 were constructed by integrating a random normal distribution 
to form Brownian noise.
Each selected noise model was normalized to the desired red noise amplitude, and 
then combined with white noise and a simulated eclipse curve to form a 
synthetic light curve.

\label{f:rnmodels}
}
\end{figure}


\begin{figure}
{\centering
 \includegraphics[width=0.95 \textwidth, angle=0]{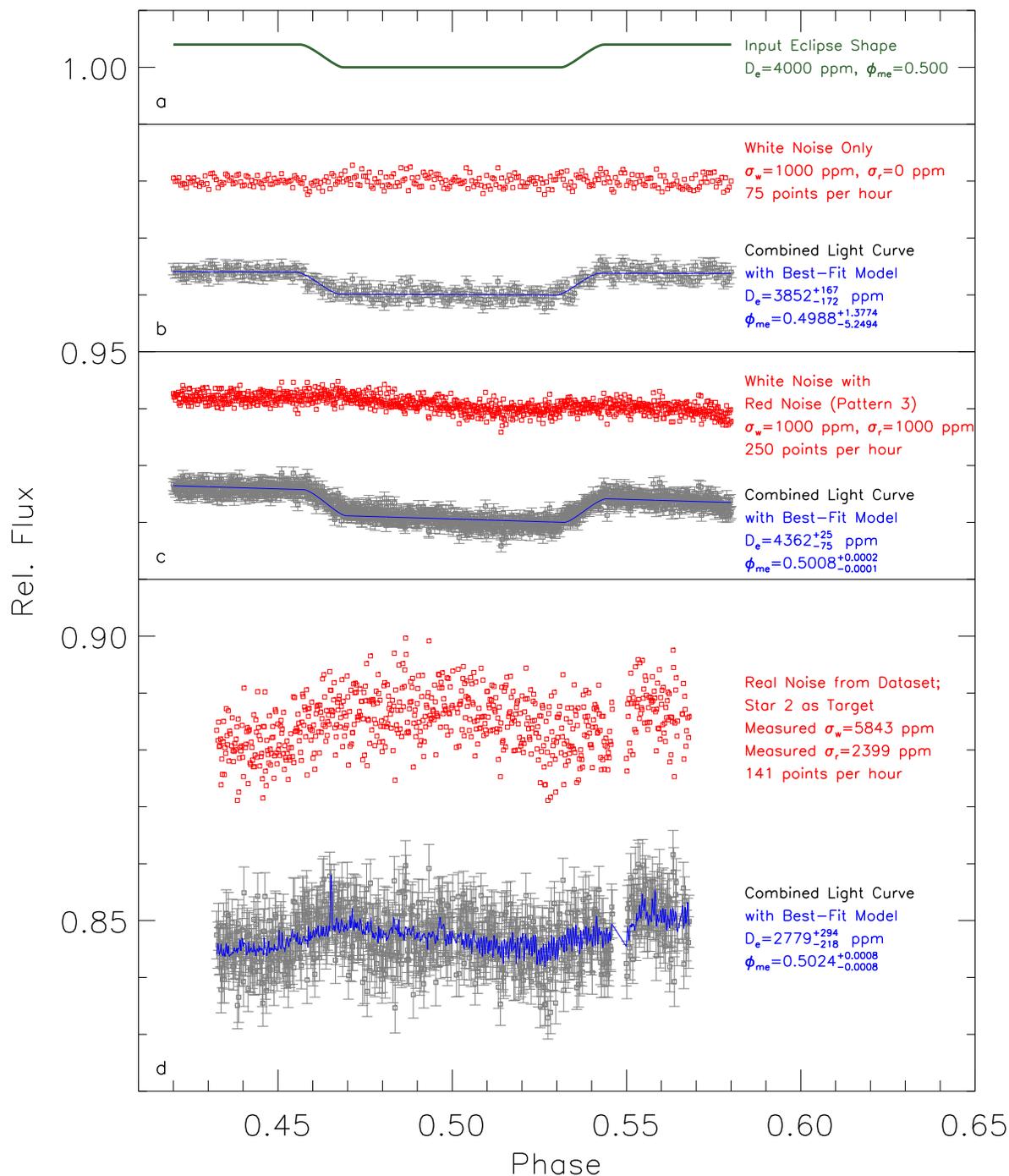}}
 \caption[]{
 Examples of construction of eclipse light curves: a 4000-ppm deep 
 eclipse with no noise at all (a), 
 and with each of the three types of noise models (shown in red):
 White noise only (b), white noise plus red noise (c), 
 and real noise systematics from NICFPS Ks-band photometry (d).
For each case, we combined the eclipse shape and noise into 
a light curve to model with the MCMC and recover the best-fit 
depth and mid-eclipse phase.
 \label{f:allkinds}
 }
 \end{figure}


\begin{figure}
{\centering
  \includegraphics[width=0.85 \textwidth,angle=0]{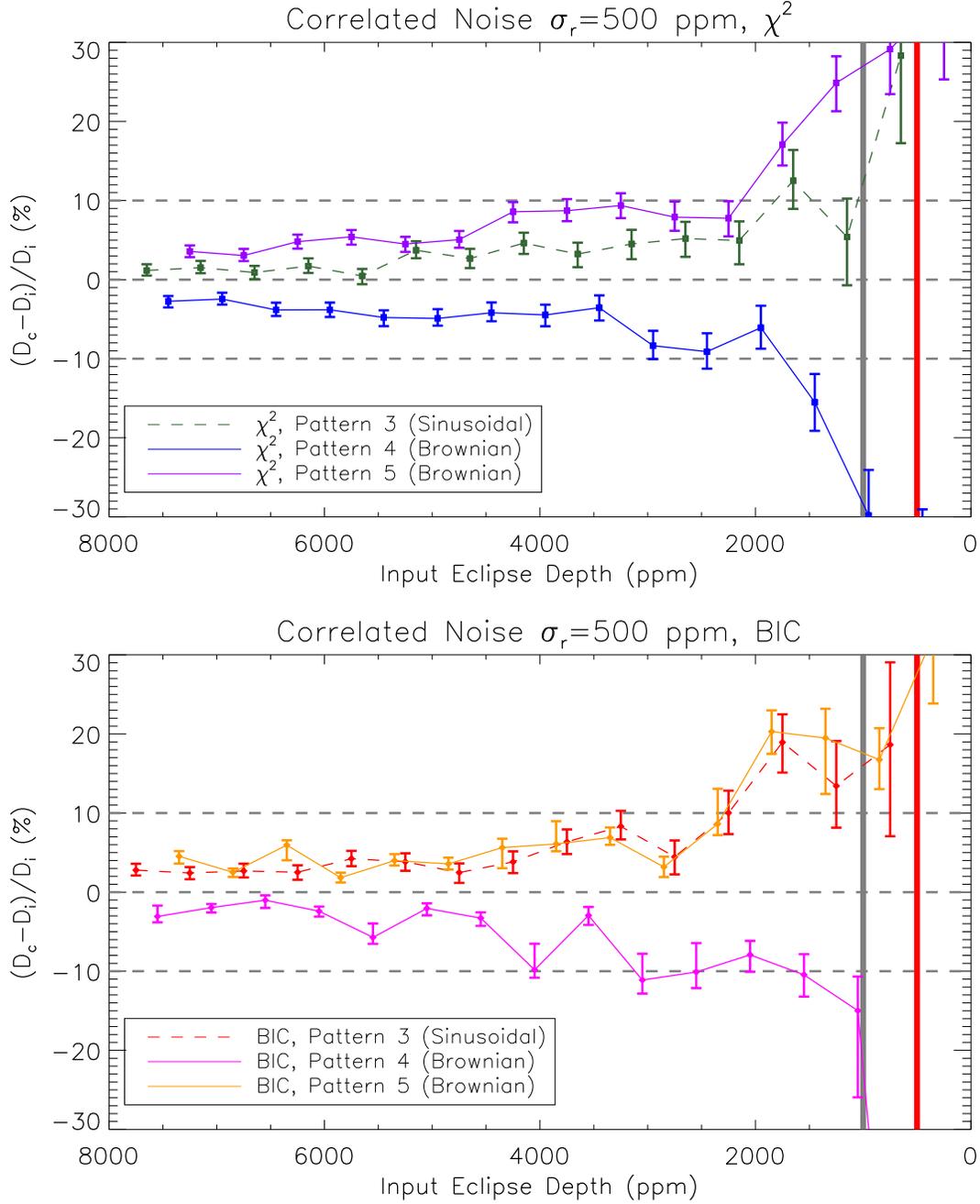}}
  \caption[]{Comparison of the two different types of red noise models we used in these analyses: 
  a sum of sinusoids (Pattern 3), and two Brownian random walks 
  (Patterns 4 and 5, see Figure~\ref{f:rnmodels}), with amplitudes 
  of $\sigma_w$=1000 ppm, $\sigma_r$=500 ppm, shown by the vertical gray and red bars.
These tests all used the 240-minute observing window with a cadence of 250 points per hour. 
The accuracy is expressed as a percent difference between the input 
and recovered depth. 
  \label{f:rn45}
  }
  \end{figure}

\appendix

\section{Benchmark Data Sets}\label{sec:benchmark} 

Given the biases and complications in detecting exoplanet eclipses, and the diversity of routines 
developed to make these measurements, there is a strong need for common
reference datasets containing realistic noise levels and patterns against which different
teams can test the performance of their routines.
Such datasets are not commonly available; thus, we provide here 
a set of models to serve this purpose for the exoplanet community 

We selected a number of representative light curves from our studies that can be used 
as benchmark testing for any routines designed to fit and measure eclipses: 
ten datasets that span the full breadth of 
our tests, both the synthetic noise models (white and red noise) and the real data 
systematics (from the CoRoT-1 Ks-band photometry). 
Various baselines from 160 to 506 minutes and sampling rates from 
25 to 250 points per hour are featured. 
The depth and central phase of the eclipses that are 
input are provided, so that any team can check the results from their analysis routine against them.

 Table~\ref{t:benchmarks} shows the characteristics of each dataset: the type of noise (synthetic or 
 from one of the real stars' differential light curves), white and red noise amplitude 
 $\sigma_w$ and $\sigma_r$ (in ppm), 
 observing window (in minutes), sampling rate (in points per hour), and the 
 input eclipse signal's depth 
 (in ppm) and mid-eclipse phase. 
 For the synthetic datasets, $\sigma_w$ and $\sigma_r$ are the input values that we started with; 
 for the datasets with real noise, these values were measured by fitting to Equation~\ref{e:rednoise}.
 These datasets are made available in machine-readable ASCII format on the 
 Vizier Data Service.\footnote{For the time being, they can be found at 
\url{http://www.pha.jhu.edu/~rogers/vhj/mcmc/benchmark/}; 
this is to be replaced later with Vizier url and citation, when the paper is accepted.}

 The first seven sets, labeled S1 through S7, are synthetic light curves constructed in the manner 
 described in Section~\ref{ss:lcconstruction}. Sets S1 through S4 contain white noise only, while S5 
 through S7 also include red noise components (S5 and S6 use Pattern 3, S7 uses Pattern 4; see Figure~\ref{f:rnmodels}). 
 Sets S1 and S6 are also the data shown in panels (b) and (c), respectively, of Figure~\ref{f:allkinds}.
 The first several lines of set S1 are shown in Table~\ref{t:synthlc} as an example 
 to illustrate the format and data types.
 The first three columns show the orbital phase at which each observation occurs, the observed flux, and 
 the flux uncertainty. 
 Only these three columns are necessary for testing purposes; the remaining four columns 
 simply separate the different components of the construction: the noiseless eclipse shape, the 
 combined noise base, and the white and red noise components.
 A greater number of significant digits are given in the online tables.
 
The three remaining datasets R1, R2, and R3 use the noise and systematics from the photometry of 
the CoRoT-1 field in Ks. 
Set R1 treats comparison star 2 as the target and adds a 4000-ppm eclipse, modeled using the characteristics of 
 ``Exoplanet X'' from Table~\ref{t:exopx}. Refer to that table and Section~\ref{ss:lcconstruction} 
 for all the other parameters that are needed 
 to reconstruct the eclipse signal in light curve.
 This is the light curve shown in panel (d) of Figure~\ref{f:allkinds}; it is also 
 equivalent to the eclipseless light curve 2-a described in Section~\ref{ss:rdataset} with a 4000-ppm 
 eclipse shape added.
 Sets R2 and R3 use the light curve 
 from the actual target star, with set R2 having an extra 1500-ppm eclipse added on top of the 
 true signal. If a routine is working properly it should find an eclipse 1500 ppm deeper for R2 than for 
 R3. Table~\ref{t:reallc} shows the first several lines of set R1. The data to fit is again in the first three 
 columns, but the real noise sets include the other variables -- offset position, $z$, $w$, $x$, $y$, and $s$ -- 
 that are needed for the full model described in Section~\ref{ss:lcmodel}.
  
In order to facilitate tests of eclipse analysis processes, understand their accuracy and limitations, 
and establish a consistent standard, we propose to use these datasets as a simple set of benchmark tests.  
Before being used to claim a secondary eclipse detection, any analysis routine should be tested on 
these or similar synthetic light curves incorporating both random and correlated noise.  
Next, whenever possible, an artificial eclipse signal should be added to other stars in the 
target dataset to test the behavior with the noise effects unique to that dataset.  
Understanding the limitations of the dataset and analysis routine will give a much better sense of 
the reliability of the resulting eclipse measurements.

\end{document}